\documentclass[fleqn,usenatbib]{mnras}

\usepackage[T1]{fontenc}
\usepackage{ae,aecompl}

\usepackage{graphicx}   
\usepackage{amsmath}    
\usepackage{amssymb}    

\usepackage{url}

\newcommand{\kms}{~km s$^{-1}$~}
\newcommand{\AG}{AG~Peg~}
\newcommand{\AGE}{AG~Peg}
\newcommand{\dotM}{~M$_{\odot}$~yr$^{-1}$~}

\newcommand{\Swift}{{\it Swift~}}
\newcommand{\SwiftE}{{\it Swift}}
\newcommand{\Rosat}{{\it ROSAT~}}
\newcommand{\RosatE}{{\it ROSAT}}
\newcommand{\xspec}{{\sc xspec~}}

\title[X-rays from \AGE]
{Recent X-ray observations of the symbiotic star \AGE:
do they signify Colliding Stellar Winds?}

\author[S.A.Zhekov and T.Tomov]{Svetozar A. Zhekov$^1$\thanks{
E-mail: szhekov@astro.bas.bg; toma.tomov@astri.umk.pl.} 
and Toma Tomov$^2$\\
$^1$Institute of Astronomy and National Astronomical Observatory, 
72 Tsarigradsko Chaussee Blvd., Sofia 1784, Bulgaria\\
$^2$Centre for Astronomy, Faculty of Physics, Astronomy and
Informatics, Nicolaus Copernicus University, Grudziadzka 5, 87-100
Torun, Poland
}

\date{}

\pubyear{2016}

\begin{document}
\label{firstpage}
\pagerange{\pageref{firstpage}--\pageref{lastpage}}
\maketitle

\begin{abstract}
We present an analysis of recent X-ray observations of the symbiotic 
star \AGE. 
The X-ray emission of \AG as observed with \Swift in 2015 shows
considerable variability on time scale of days as variability on 
shorter time scales might be present as well.
Analysis of the X-ray spectra obtained in 2013 and 2015 confirms
that \AG is an X-ray source of class $\beta$ of the X-ray sources
amongst the symbiotic stars.
The X-ray emission of \AG as observed with \Rosat (1993 June) 
might  well originate 
from colliding stellar winds (CSW) in binary system.
On the other hand, the characteristics of the X-ray emission of
\AG in 2013 and 2015 (\Swift) are hard to accommodate 
in the framework of the CSW picture.
Analysis of the light curves in 2015 shows that the power
spectrum of the X-ray variability in \AG resembles that 
of the flicker noise (or flickering) being typical for
accretion processes in astronomical objects.
This is a sign that CSWs did not play a key
role for the X-ray emission from \AG  in 2013-2015 and a
different mechanism (probably accretion) is 
also getting into play.

\end{abstract}




\section{Introduction}
\AG (HD 207757) is an old symbiotic nova whose outburst occurred in 
the mid-19th century and was characterized with a slow 
rise to its maximum and a slower decline (lasting for many decades). 
This is likely the slowest classical nova eruption amongst
recorded until present day (e.g., \citealt{kenyon_93}; 
\citealt*{kenyon_01} and references therein).
Recently, there were reports about ongoing changes of the \AG emission
in the optical 
\citep{munari_13} and in the X-rays (\citealt{nunez_13}; 
\citealt{luna_15}; \citealt{ramsay_15}). This gives solid indications 
about a new type of activity in the evolution of this old symbiotic 
nova.

\AG was first detected in X-rays with \Rosat \citep*{murset_95}. 
It is an X-ray source that falls in the class $\beta$ of the
X-ray sources amongst symbiotic stars: these are sources that show
emission from an optically thin plasma with a temperature of a few
10$^6$~K \citep*{murset_97}.

The X-ray emission of the $\beta$-class sources is believed to
originate from colliding stellar winds (CSW) in binary system (e.g.,
\citealt{murset_97}; \citealt{luna_13}). Since a symbiotic binary
consists of a red giant (possessing a massive but slow wind) and a 
white dwarf, that could well be the case provided the latter component 
has a stellar wind as well. This was exactly the case for \AG as shown 
by analysis of the Hubble Space Telescope spectra \citep*{nuss_95}.
However, the reported recent activity of \AG indicates that the
physical picture in this old symbiotic nova might be more complicated
than that.

In this work, our goal is to address the origin of the X-ray emission
from the symbiotic star \AG by making use of the available X-ray data.
In Section ~\ref{sec:data}, we review the X-ray observations of \AGE.
In Section ~\ref{sec:results}, we present results from analysis of the
X-ray properties of \AGE. In Section ~\ref{sec:xray_origin}, we 
discuss the origin of the X-ray emission of this old symbiotic nova. 
Our conclusions are listed in Section ~\ref{sec:conclusions}.

\begin{figure*}
\begin{center}
\includegraphics[width=2.8in, height=2.0in]{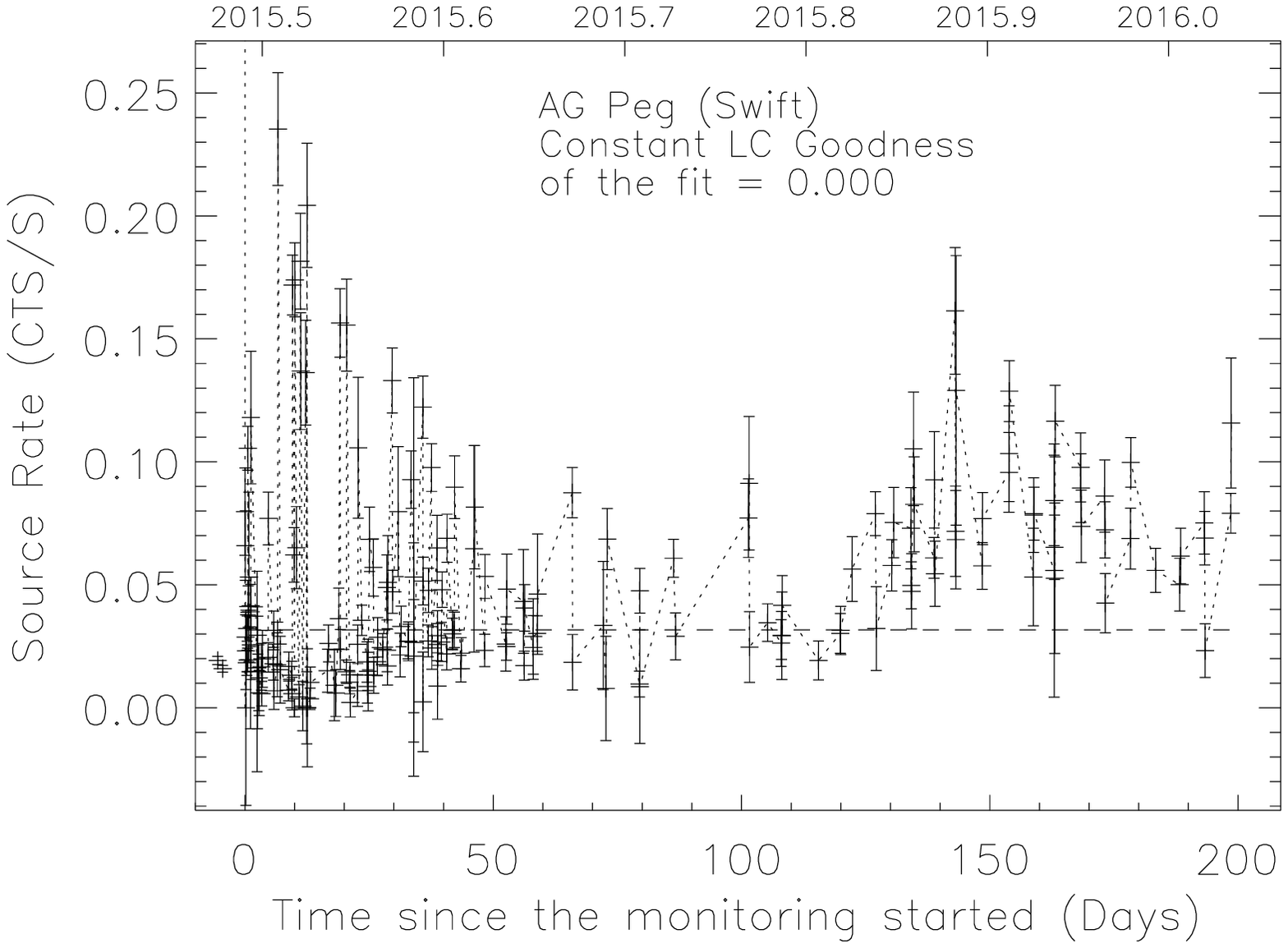}
\includegraphics[width=2.8in, height=2.0in]{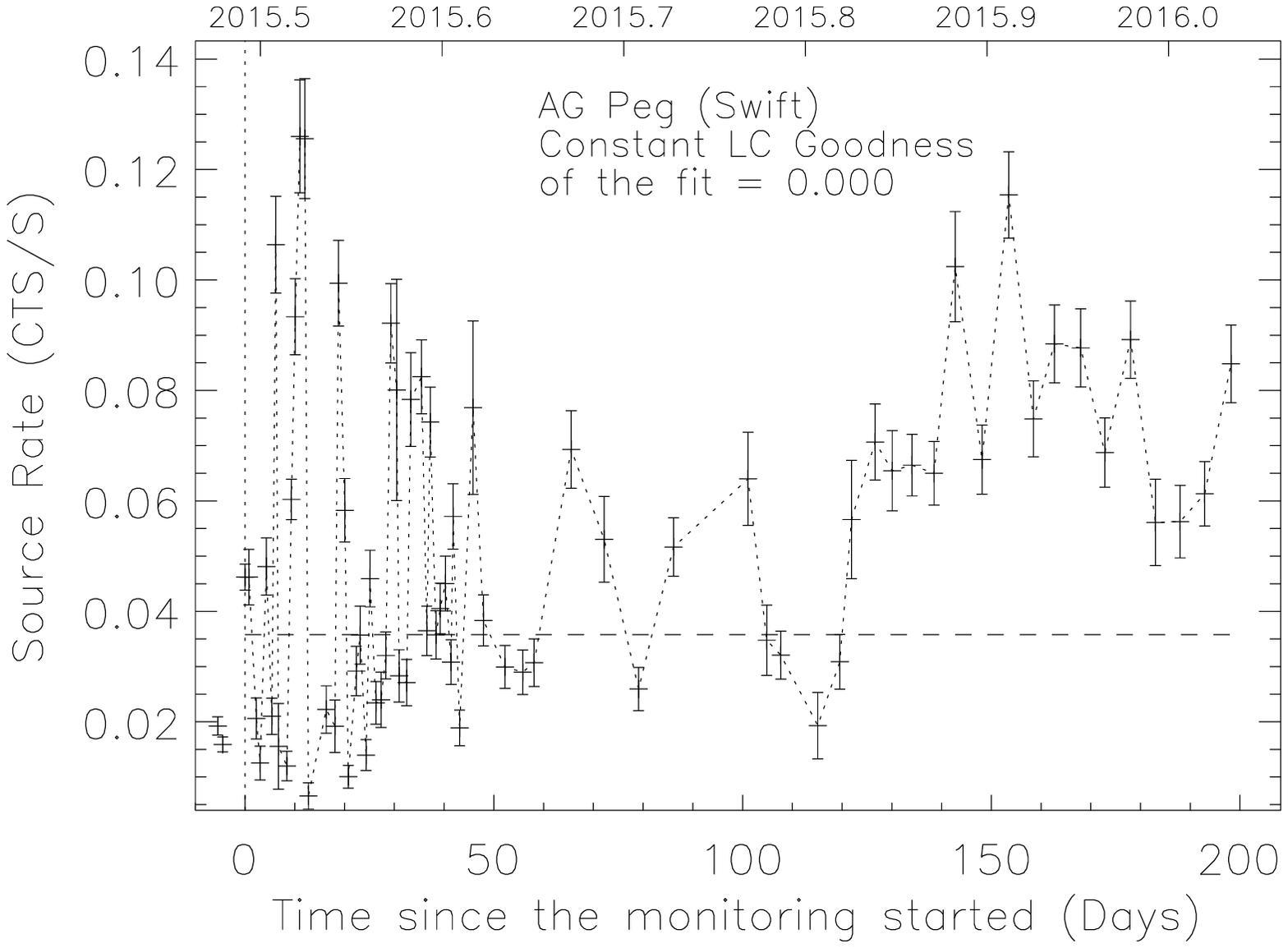}
\end{center}
\caption{The (0.3 - 3 keV) background-subtracted light curves
(LC) of \AG in 2015 (\SwiftE).
{\it Left} panel: the `GTI' LC based on all the 71 observations.
{\it Right} panel: the `daily' LC based on the average count rate
for each data set.
The corresponding constant flux level is denoted by a dashed line.
The count rates of both observations in 2013 are given for
comparison at some fiducial negative time.
}
\label{fig:totalLC}
\end{figure*}

\section{X-ray data}
\label{sec:data}
In this study, we used archive data of \AG from the X-ray
observatories \Rosat and \SwiftE.

\RosatE.
We used one pointed observation taken on 1993 June 9 (ObsID
rp300186n00). 
Following the recommendations for the \Rosat Data
Processing\footnote{\url{http:
//heasarc.gsfc.nasa.gov/docs/rosat/rhp_proc_analysis.html}}, we
extracted the source and background spectra. Since the data were 
taken after 1991 Oct 14, we adopted the response matrix 
pspcb\_gain2\_256.rmf and we used the package {\it pcarf} to 
construct the ancillary response file.
The extracted spectrum has 409 net source counts in an
effective exposure of 5788 s.

\SwiftE.
We used 73 pointed observations (ObsID from
00032906001 to 00032906079; data sets with ObsID with the last two 
digits 17, 31, 38, 40, 49 and 51 do not exist in the archive) 
taken in 2013 August (two data sets) and in 2015 June - 2016 January
(71 data sets). 
Following the \Swift XRT Data Reduction
Guide\footnote{\url{http://swift.gsfc.nasa.gov/analysis/xrt_swguide_v1_2.pdf}},
we extracted the source and background spectra for each observation.
Extraction regions had the same shape and size for each data set.
For our analysis, we used the response matrix 
(swxpc0to12s6\_20010101v014.rmf)  provided by
the most recent (2015 July 31) \Swift calibration
files\footnote{\url{http://heasarc.gsfc.nasa.gov/docs/heasarc/caldb/swift/}}
and we also used the package {\it xrtmkarf} to construct the ancillary 
response file for each data set.
Due to the relatively short exposure times (e.g., between 200 and
9,000 s), the individual spectra have limited photon statistics  in
the (0.3 - 3 keV) energy band: the net source counts are between 
{\it even} less than 10 and $\sim 300$ (`typically' a few tens 
of counts). The total number of source counts in the entire sample of
\Swift spectra in 2015 is 6,440 (71 spectra) and 299 in 2013 (2
spectra). And, for those with more than 100 source counts (29 spectra), 
we have 4,586 net source counts in total.

For the spectral analysis in this study, we made use of standard as
well as custom models in version 11.3.2 of \xspec 
\citep{Arnaud96}.

\section{Results}
\label{sec:results}
The available data allows us to obtain some pieces of information on 
the X-ray properties of \AGE: e.g., about its X-ray variability and 
the characteristics of its X-ray spectrum. 

\subsection{X-ray light curve}
\label{sec:xray_lc}
It is worth noting that the good time intervals (GTI) in the \Swift
observations of \AG are unevenly spaced. As a rule of thumb, the
length of a GTI segment is shorter than the time `gap' between two
consecutive GTIs. This prevents constructing a `standard' light curve 
(LC), that is a LC with equal time bins. For that reason, we
constructed the following background-subtracted LCs: one, `GTI' LC, 
that includes all the count rates from individual GTIs from all the 
observations in 2015 (195 GTIs in total) and another one, `daily' LC, 
that includes all the average count rates for each data set (71 data 
sets). These LCs of \AG are shown in Fig.~\ref{fig:totalLC}. 
Examples of
the individual LCs (both from \Swift and \RosatE), namely,
those including only the GTIs for that specific date of observation, 
are shown in Appendix~\ref{append:X-ray} (see Fig.~\ref{fig:dailyLC1}).

To address the X-ray variability of \AGE, we fitted a constant count
rate model to the `daily' as well as the `GTI' LCs. The formal
goodness of fit, using $\chi^2$ fitting,  is very small (equal to 
zero) for both LCs. This undoubtedly shows that \AG is a variable 
X-ray source on time scales of days: the \Swift observations in 2015 
span a time interval of $\sim 200$ days.
Applying the Lomb-Scargle method (\citealt{lomb_76}; 
\citealt{scargle_82}; see \S~13.8 in \citealt{press_92}), 
no periodicity  was revealed in either of the X-ray LCs.

On the other hand, the individual LCs show both cases, that is of 
constant and variable flux, which means that variable X-ray emission 
on time scales of a few hours is not always present in \AG
(note that only LCs with the goodness of the fit $< 0.001$ were assumed
variable and such were $\sim 25$\% of all the idividual LCs). 
However, 
due to the limited photon statistics of the \Swift data on \AG 
caution is needed before drawing a firm conclusion. 
Future uninterrupted observations with higher sensitivity will be 
crucial to study the short-time ($< 1$ day) X-ray variability of 
this object.

\begin{figure}
\begin{center}
\includegraphics[width=2.8in, height=2.0in]{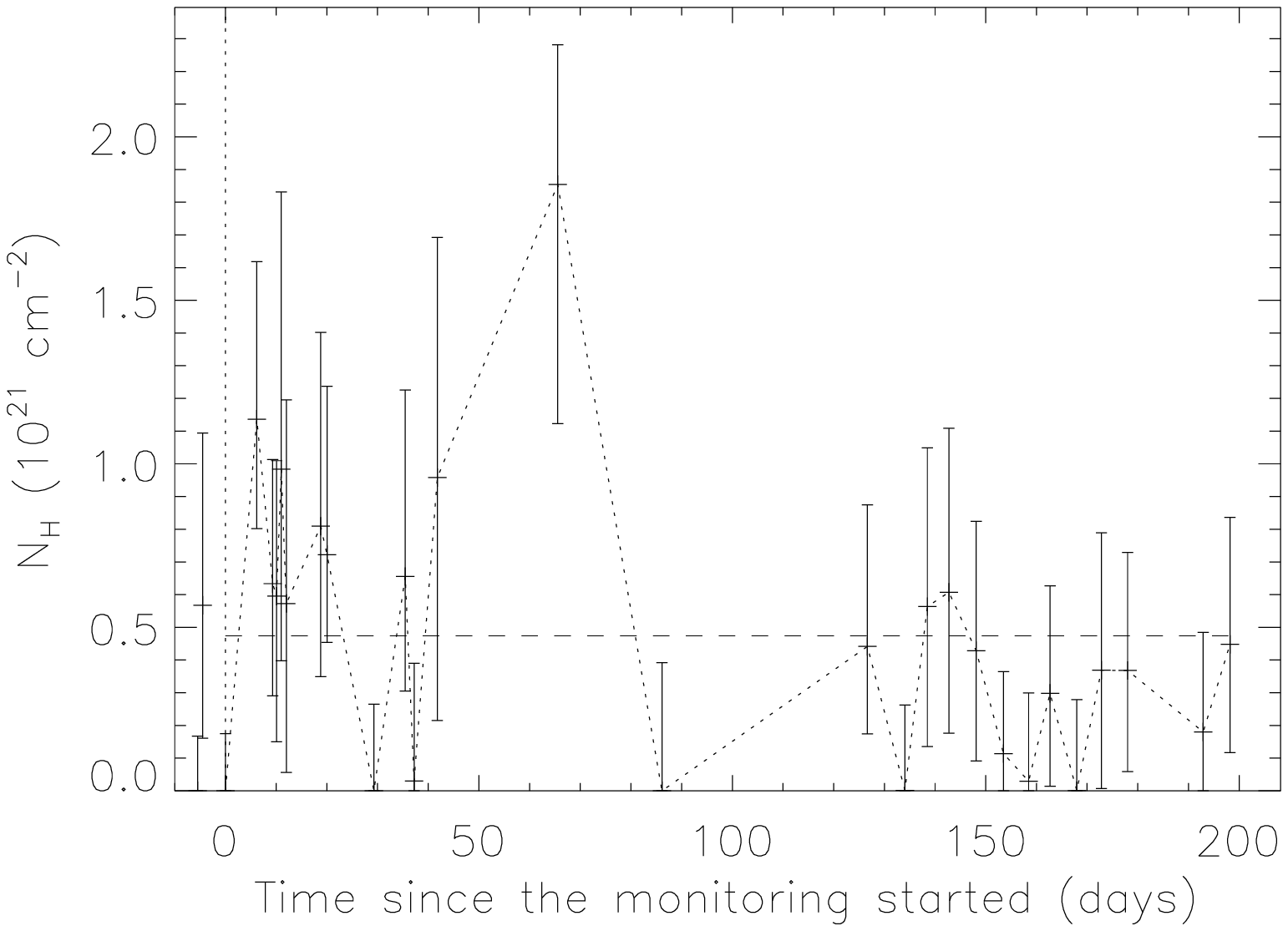}
\includegraphics[width=2.8in, height=2.0in]{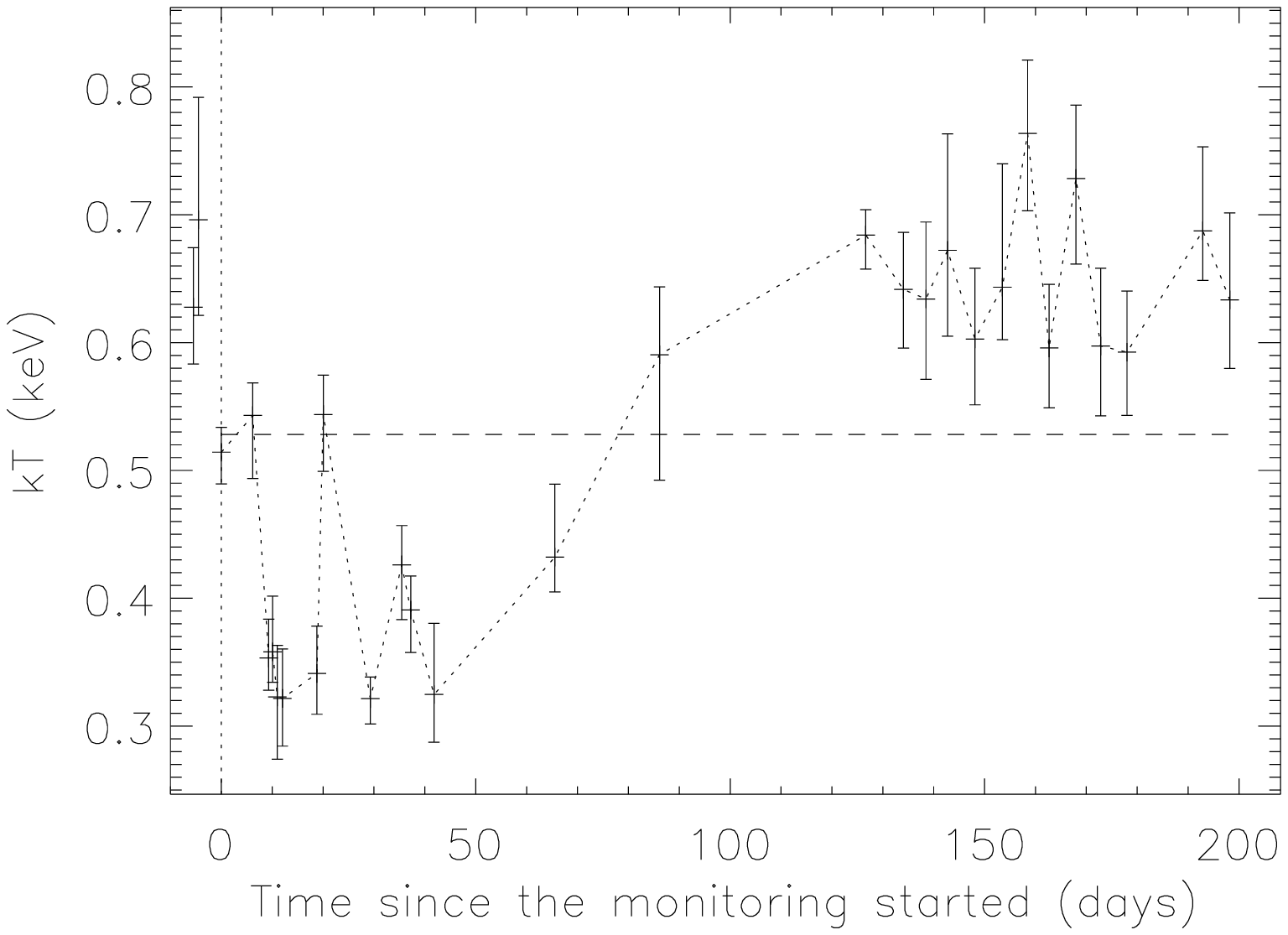}
\includegraphics[width=2.8in, height=2.0in]{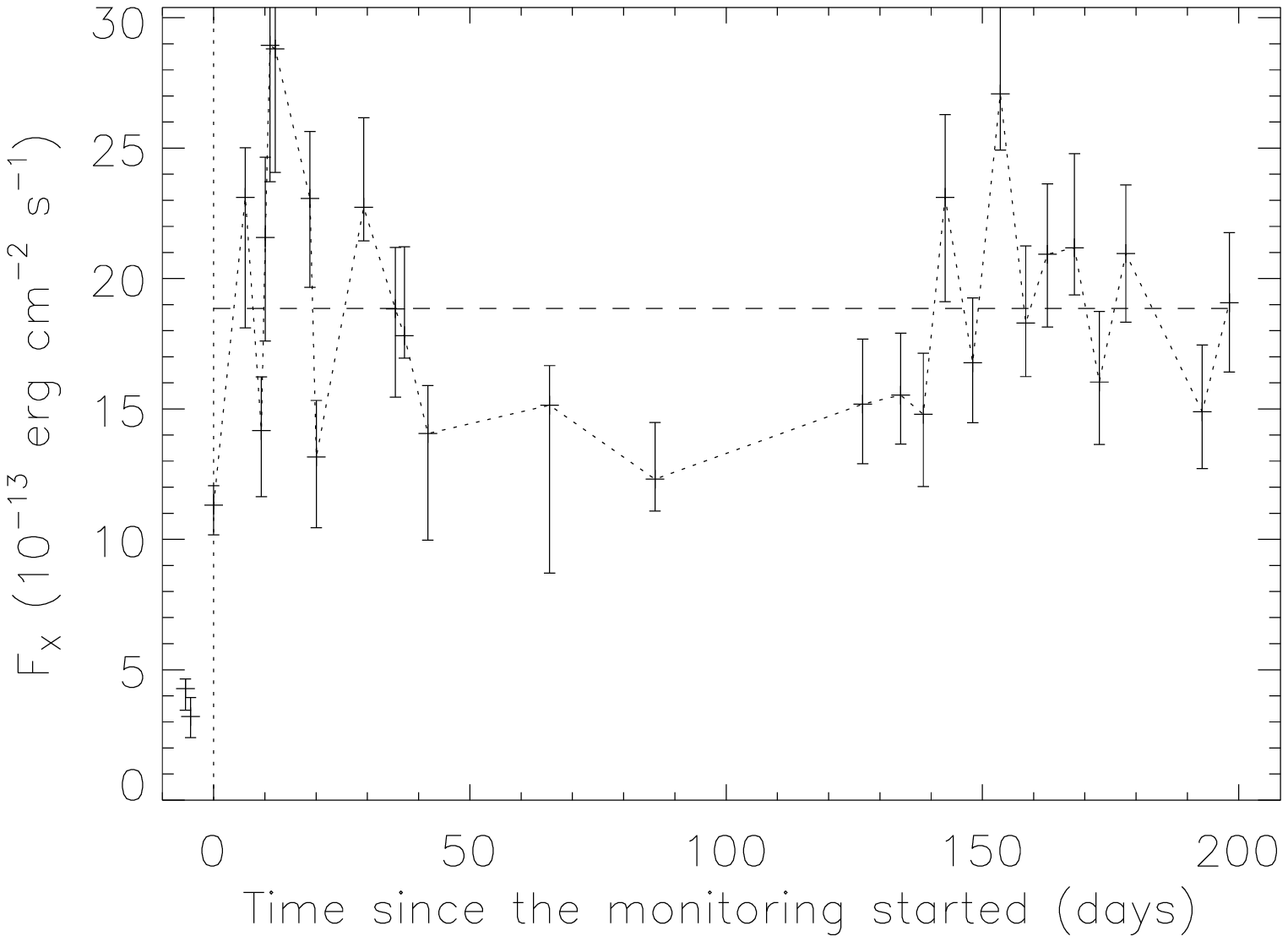}
\end{center}
\caption{Results from the fits to the \Swift spectra of \AG with 
optically thin plasma model for the data sets with higher than 
100 X-ray counts.
Shown are the time evolution of 
the X-ray absorption (N$_H$), plasma temperature 
(kT; 1 keV $= 11.6\times10^6$ K) and 
observed X-ray flux (0.3 - 3 keV). The horizontal dashed line 
indicates the mean value for the corresponding parameter.
The results for both observations in 2013 are given 
at some fiducial negative time.
}
\label{fig:swift_1T}
\end{figure}

\begin{table*}
\centering
\caption{Spectral Model Results (\SwiftE)
\label{tab:swift}}
\begin{tabular}{llllll}
\hline
\multicolumn{1}{c}{Model case} & 
   \multicolumn{1}{c}{N$_H$} & \multicolumn{1}{c}{kT} &
   \multicolumn{1}{c}{norm} & \multicolumn{1}{c}{F$_X$} &
   \multicolumn{1}{c}{L$_X$} \\
\hline
 1T      & 0.69 (1.29) & 0.59 (0.30) & 
           1.67 (1.13) $\times 10^{-3}$ & 12.2 (7.0) & 
           13.6 (9.5) \\ 
 1T (\small N$_H$)  & 0.32 & 0.64 (0.35) & 
               1.41 (0.83) $\times 10^{-3}$ & 12.2 (7.1) & 
               11.4 (6.8) \\ 
 1T (\small kT)  & 1.43 (1.10) & 0.45 & 
            3.07 (2.06) $\times 10^{-3}$ & 12.2 (7.0) & 
            19.8 (12.5) \\ 
 1T (\small N$_H$, kT) & 0.35 & 0.52 & 
                  1.65 (0.97) $\times 10^{-3}$ & 12.0 (7.0) & 
                  11.7 (6.7) \\ 
 CSW  & 1.58 (1.07) & \dots & 3.22 (2.15) & 12.3 (7.2) & 
        23.3 (14.9) \\ 
 CSW (\small N$_H$) & 0.79 & \dots & 1.97 (1.13) & 12.3 (7.0) & 
                      15.8 (9.1) \\ 

\hline

\end{tabular}

Note --
Results from the fits to the \Swift spectra of \AGE.
The `1T' and `CSW' notations  indicate the cases of one-temperature
thin-plasma and colliding stellar wind models, respectively. The model
parameter common (being the same) for all the spectra is given in
parentheses therein. The model parameters are
the X-ray absorption N$_H$ in units of 10$^{21}$ cm$^{-2}$, 
the plasma temperature kT in keV;
the \xspec model normalization (norm), 
the observed flux (0.3 - 3 keV energy range) F$_X$ in units of 
$10^{-13}$ ergs cm$^{-2}$ s$^{-1}$ and
luminosity in units of $10^{31}$ ergs s$^{-1}$ for adopted distance 
of 800 pc (\citealt{kenyon_93}, \citealt{kenyon_01}).
For each of them, given are the mean value followed in parentheses
by the standard deviation for the entire sample of 73 spectra. 

\end{table*}

\begin{figure*}
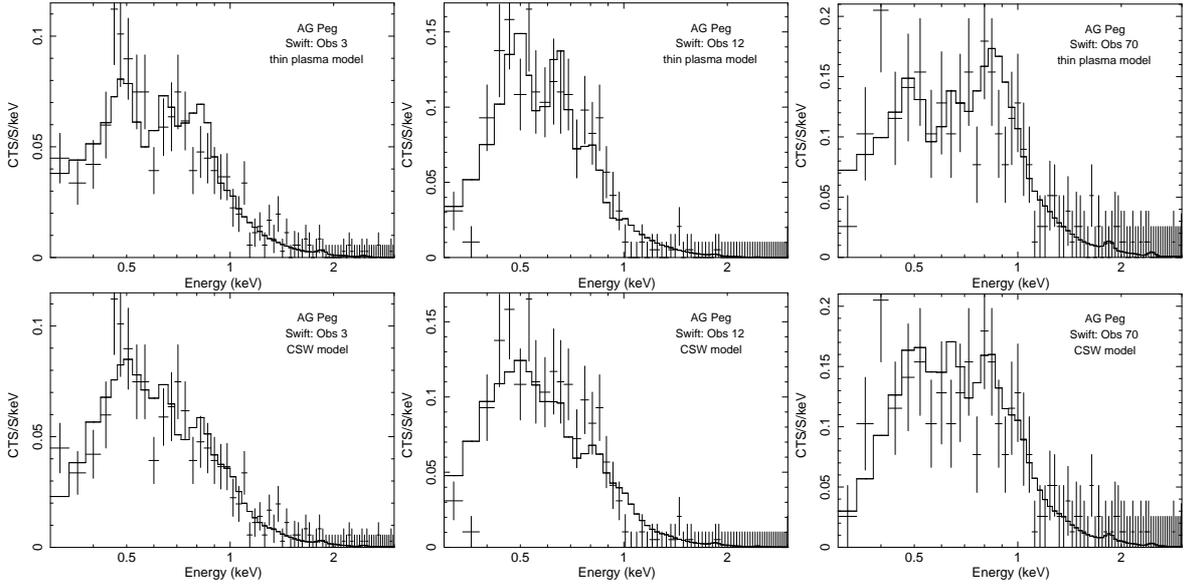

\begin{center}
\includegraphics[width=1.5in, height=2.0in, angle=-90]{fig3a.eps}
\includegraphics[width=1.5in, height=2.0in, angle=-90]{fig3b.eps}
\includegraphics[width=1.5in, height=2.0in, angle=-90]{fig3c.eps}
\includegraphics[width=1.5in, height=2.0in, angle=-90]{fig3d.eps}
\includegraphics[width=1.5in, height=2.0in, angle=-90]{fig3e.eps}
\includegraphics[width=1.5in, height=2.0in, angle=-90]{fig3f.eps}
\end{center}
\caption{The \Swift spectra of \AG that have more than 200 X-ray
counts overlaid  with the optically-thin plasma model fit (upper row)
and with the CSW model fit for the case of CIE plasma (lower row). The
spectra were slightly re-binned for presentation.
}
\label{fig:spec_swift}
\end{figure*}

\subsection{X-ray spectrum}
\label{sec:xray_spec}
We recall that the \Swift data on \AG have limited photon statistics
(see Section~\ref{sec:data}). A standard way to analyse such data is 
to construct a total spectrum which could have much higher quality.
However, 
because of 
the high level of variability detected in the X-ray emission 
from this object (see Section~\ref{sec:xray_lc}), we instead analysed 
the individual spectra aimed at studying
the global spectral properties of the X-ray 
emission from \AG and to address how they change in time.

Namely, we made use of the implementation of the Cash statistic 
\citep{cash_79} in \xspec for the model fits
to the unbinned \Swift spectra in the (0.3 - 3 keV) energy range. We 
note that the X-ray background is very low: its mean contribution in
our sample of spectra is 4\% of the total counts in extracted 
source spectra. Thus, we could
simply neglect its contribution. Nevertheless, we constructed a common
(total) background spectrum of all the \Swift observations of \AGE. A
power-law model provided a very good fit to this total background
spectrum, So, such a spectral component with fixed model parameters was 
used in the further fitting of the individual spectra.

\subsubsection{Discrete-temperature models}
\label{sec:1T_model}
Since \AG falls in the class $\beta$ of the X-ray sources amongst 
symbiotic stars (showing emission form an optically thin plasma 
with a temperature of a few 10$^6$~K; \citealt{murset_97}), the
adopted spectral model was a sum of an absorbed optically thin plasma
emission and a background component, as explained above.
For the H, He, C and N abundances, we adopted the values derived by
\citet{schmutz_96}: H : He : C : N = 1 : 0.1 : $10^{-5}$ : $10^{-3}$ 
by number (H $= 1$, He $= 1.02$, C = $0.03$, N $= 8.93$ with respect 
to the solar abundances; \citealt{ag_89}). 
And to improve the quality of the fits, the O, Ne, Mg, Si and 
Fe abundances were allowed to vary. It is important to note that 
the spectral fitting was performed in two steps. First, we used only
those spectra that have more than 100 X-ray counts. These spectra (29
in total) were fitted simultaneously allowing the O, Ne, Mg, Si and Fe
abundances to vary. 
The derived abundance values (with $1\sigma$ errors) are:
O $= 0.67^{+0.05}_{-0.04} $,
Ne $= 0.00^{+0.02}_{-0.00} $, 
Mg $= 0.02^{+0.05}_{-0.02} $, 
Si $= 0.32^{+0.09}_{-0.05} $, 
Fe $= 0.16^{+0.01}_{-0.01} $
~(the values are with respect to the solar abundances; 
\citealt{ag_89}). 
Second, each individual spectrum was fitted
with the same spectral model having abundance values fixed to
those derived in the first step of the fitting. The free
parameters for the individual fits were the amount of X-ray
absorption, plasma temperature and normalization parameter. 

Some fit results are shown in Figs.~\ref{fig:swift_1T} and
~\ref{fig:spec_swift}  as the formal 
goodness of the fit is very good for all the data 
sets. We note that there might be some trend in the change (increase)
with time of the temperature of the X-ray emitting plasma as indicated 
by the results from the fit to the spectra with the better quality 
(X-ray counts higher than 100). The linear Pearson correlation 
coefficient is 0.83. Also, the X-ray absorption could be decreasing 
with time but the correlation is weak (the correlation coefficient is 
-0.38).

For completeness, we explored also the following cases of 
discrete-temperature models. We considered  absorbed
one-temperature thin-plasma emission as above but assuming that all
the spectra were subject to the same X-ray absorption. Next, we
adopted  one-temperature plasma model but assuming that the
plasma temperature was not changing with time. Finally, we considered
the case where neither the X-ray absorption nor the plasma temperature
were changing with time. In all these case, the spectral fits were
performed following the same two-step procedure as explained above. We
note that the quality of the fits was acceptable in all the cases
under consideration, slightly decreasing with increasing the number of
model parameters common for all the data sets: e.g., sharing the same
X-ray absorption, or the same plasma temperature, or both. 

Some basic results from the discrete-temperature model fits are
summarized in Table~\ref{tab:swift}. We see that it is safe to
conclude that considerable variability of the basic parameters of the
X-ray emission from \AG is present. This is valid for all the cases
of absorbed one-temperature thin-plasma emission considered here.
Namely, the value of the standard deviation for a given parameter is 
always more than 50\% of the mean value of that model parameter for
the sample of 73 \Swift spectra at hand. This is indicative of a
considerable scatter of that parameter, that is of its considerable
variability with time.
This conclusion is also supported by the large $\chi^2$-value for 
the sample of derived values of a given parameter with respect of 
their associated mean, which results in a very small probability 
(equal to zero) for constancy of that parameter.

\subsubsection{CSW model spectra}
\label{sec:csw_model}
We recall that from the analysis of the \Rosat observation of \AG  
\citet{murset_95} proposed that the X-ray emission of this
symbiotic star most likely originates from colliding stellar winds.
They also provided comparison of some physical quantities that
were derived in hydrodynamic calculations with those deduced in the
analysis of the \Rosat data. However, they did not provide a direct 
comparison of the theoretical and observed X-ray spectra. Here, we
attempted such a direct comparison although in some approximate
manner.

We note that the basic input parameters for the hydrodynamic model
in CSW binaries are the mass loss and velocity of the stellar winds 
of the binary components and the binary separation. The former define
a dimensionless parameter 
$\Lambda = (\dot{M}_{HS} V_{HS}) / (\dot{M}_{CS} V_{CS})$ (the
notations $HS$ and $CS$ stand for the hot star and cool star,
respectively) which
determines the shape and the structure of the CSW interaction region 
(\citealt*{luo_90}; \citealt*{stevens_92}; \citealt{mzh_93}).
We adopted the same wind parameters as in \citet{murset_95}:
$\dot{M}_{HS} = 2\times10^{-7}$\dotM, $V_{HS} = 1000$\kms,
$\dot{M}_{CS} = 4\times10^{-7}$\dotM, $V_{CS} = 20$\kms. For a binary
period of 818.2 days \citep{fekel_00} and assuming a total binary
mass of 2 solar masses, Kepler's third law gives a binary separation
of $3.22\times10^{13}$~cm. 

It is worth noting that due to its high velocity only the shocked
hot-star wind could be a source of X-ray emission. From the HS wind
parameters, one sees that the shocked HS plasma will be adiabatic.
This is indicated by the values of either of dimensionless parameters 
$\chi$ \citep{stevens_92} and $\Gamma_{ff}$ \citep{mzh_93}: 
$\chi = 16.1$ ($\chi > 1$ - adiabatic case),
$\Gamma_{ff} = 0.001$ ($\Gamma_{ff} > 1$ - cooling is important).
In general, partial electron heating might occur behind strong shocks.
However, the value of the dimensionless parameter $\Gamma_{eq} = 23.5$ 
indicates that the temperature equalization is very fast in the case 
of \AG ($\Gamma_{eq} < 1$ if the difference of electron and ion 
temperatures is important; see \citealt{zhsk_00}). 
Also, the value of the dimensionless parameter $\Gamma_{NEI} = 12.72$
indicates that the non-equilibrium ionization effects (NEI) do not
play an important role in the CSW region of \AG (the NEI effects must
be taken into account if $\Gamma_{NEI} \leq 1$ but can be neglected if
$\Gamma_{NEI} \gg 1$; see \citealt{zh_07}).

Finally, we note that the shocked cool-star plasma in the CSW region
in \AG will not be adiabatic. But, due to the low CS wind velocity
(that is, low shock velocity, correspondingly low plasma temperature) 
there will be no X-ray emission from the shock CS-wind plasma. Also, 
even if that part of the CSW region `collapses' the shock surface will 
simply `coincide' with that of the contact discontinuity (as the 
latter is calculated in the adiabatic case), which will cause no 
changes in the plasma distribution in the shocked HS-wind part of the 
CSW region.

In such an approximation, we ran a series of CSW models that consider
plasma in collisional equilibrium ionization (CIE). We used the model 
by \citet{zh_07} that is based on the 2D hydrodynamic model 
of adiabatic CSWs by \citet{mzh_93}. The latter assumes
spherical symmetry of the stellar winds that have reached their
terminal velocities before they collide. 
For the H, He, C and N abundances, we adopted the values derived by
\citet{schmutz_96}: H : He : C : N = 1 : 0.1 : $10^{-3}$ : $10^{-5}$ 
by number (H $= 1$, He $= 1.02$, C = $0.03$, N $= 8.93$ with respect 
to the solar abundances; \citealt{ag_89}).

\begin{figure}
\begin{center}
\includegraphics[width=2.0in, height=2.8in, angle=-90]{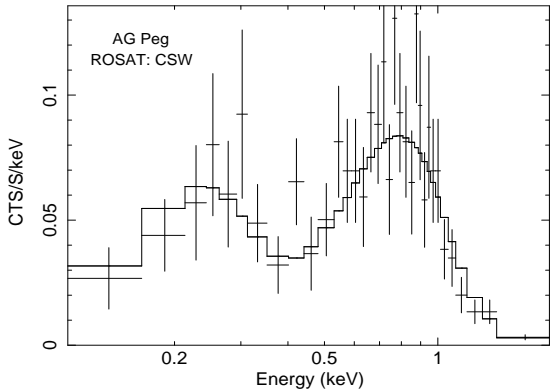}
\end{center}
\caption{The background-subtracted spectrum of \AG from \Rosat with 
the CSW model fit for the case of CIE plasma. The spectrum was 
re-binned to have a minimum of 10 counts per bin.
}
\label{fig:spec_rosat}
\end{figure}

\RosatE.
The CSW model with nominal stellar wind parameters matched well the
shape of the observed spectrum. However, the theoretical flux (or 
emission measure) was $1.36$ times higher than that observed. We 
recall that the X-ray luminosity (and flux) of the shocked 
plasma in the CSW region is $L_X \propto \dot{M}^2 V^{-3} a^{-1}$
(for the emission measure EM $\propto \dot{M}^2 V^{-2} a^{-1}$), 
where $\dot{M}$ is the stellar wind mass-loss, $V$ is the wind 
velocity and $a$ is the binary separation; see \citealt{luo_90}; 
\citealt{mzh_93}). So, CSW model with slightly reduced
mass-loss rates (0.86 of the nominal) was adopted. 

Some CSW model fit results are given in Table~\ref{tab:csw_rosat} and 
in Fig.~\ref{fig:spec_rosat}. It is worth noting that the derived 
value of the hydrogen absorption column density translates into a 
range of optical extinction A$_{\mbox{V}} = 0.14 - 0.18$ mag. 
The range corresponds to the conversion that is used:
N$_H = 2.22\times10^{21}$A$_{\mbox{V}}$~cm$^{-2}$ \citep{go_75}.
and 
N$_H = (1.6-1.7)\times10^{21}$A$_{\mbox{V}}$~cm$^{-2}$
(\citealt{vuong_03}, \citealt{getman_05});
So, we see no indication for any X-ray absorption in excess to the
optical extinction to \AG 
(E$_{\mbox{B-V}} = 0.1\pm0.05$ mag; \citealt{kenyon_93};
E$_{\mbox{B-V}} = 0.09\pm0.04$ mag; \citealt{vogel_94}).

Thus, it seems conclusive that the CSW model is capable of explaining
the X-ray emission of \AG as observed with \Rosat in 1993 June.

\begin{table}
\centering
\caption{CSW Spectral Model Results (\RosatE)
\label{tab:csw_rosat}}
\begin{tabular}{ll}
\hline
\multicolumn{1}{c}{Parameter} & \multicolumn{1}{c}{CIE} 
   \\
\hline
$\chi^2$/dof  & 23/35 \\
N$_{H}$ (10$^{21}$ cm$^{-2}$)  & 0.30$^{+0.03}_{-0.03}$ \\
$norm$  &  1.00$^{+0.07}_{-0.07}$ \\
F$_{X,1}$ ($10^{-13}$ ergs cm$^{-2}$ s$^{-1}$)  & 6.84 (11.0) \\
F$_{X,2}$ ($10^{-13}$ ergs cm$^{-2}$ s$^{-1}$)  & 6.69 (8.15) \\
\hline

\end{tabular}

Note --
Results from the fit to the \Rosat
spectrum of \AG using model spectra from the CSW hydrodynamic 
simulations for the case with collisional ionization equilibrium 
(CIE).
Tabulated quantities are the neutral hydrogen absorption column
density (N$_{H}$), the normalization parameter 
($norm$) and the absorbed X-ray flux (F$_{X,1}$ in the 0.1 - 2 keV 
energy range; F$_{X,2}$ in the 0.3 - 3 keV energy range ) 
followed in parentheses by the unabsorbed value. The $norm$ 
parameter is a dimensionless quantity that gives the ratio of 
the emission measure required by observations to that predicted by the
model.  A value of $norm = 1.0$ indicates a perfect match between both 
of them. 
The adopted abundances are those from \citet{schmutz_96}.
Errors are the $1\sigma$ values from the fit.

\end{table}

\begin{figure}
\begin{center}
\includegraphics[width=2.8in, height=2.0in]{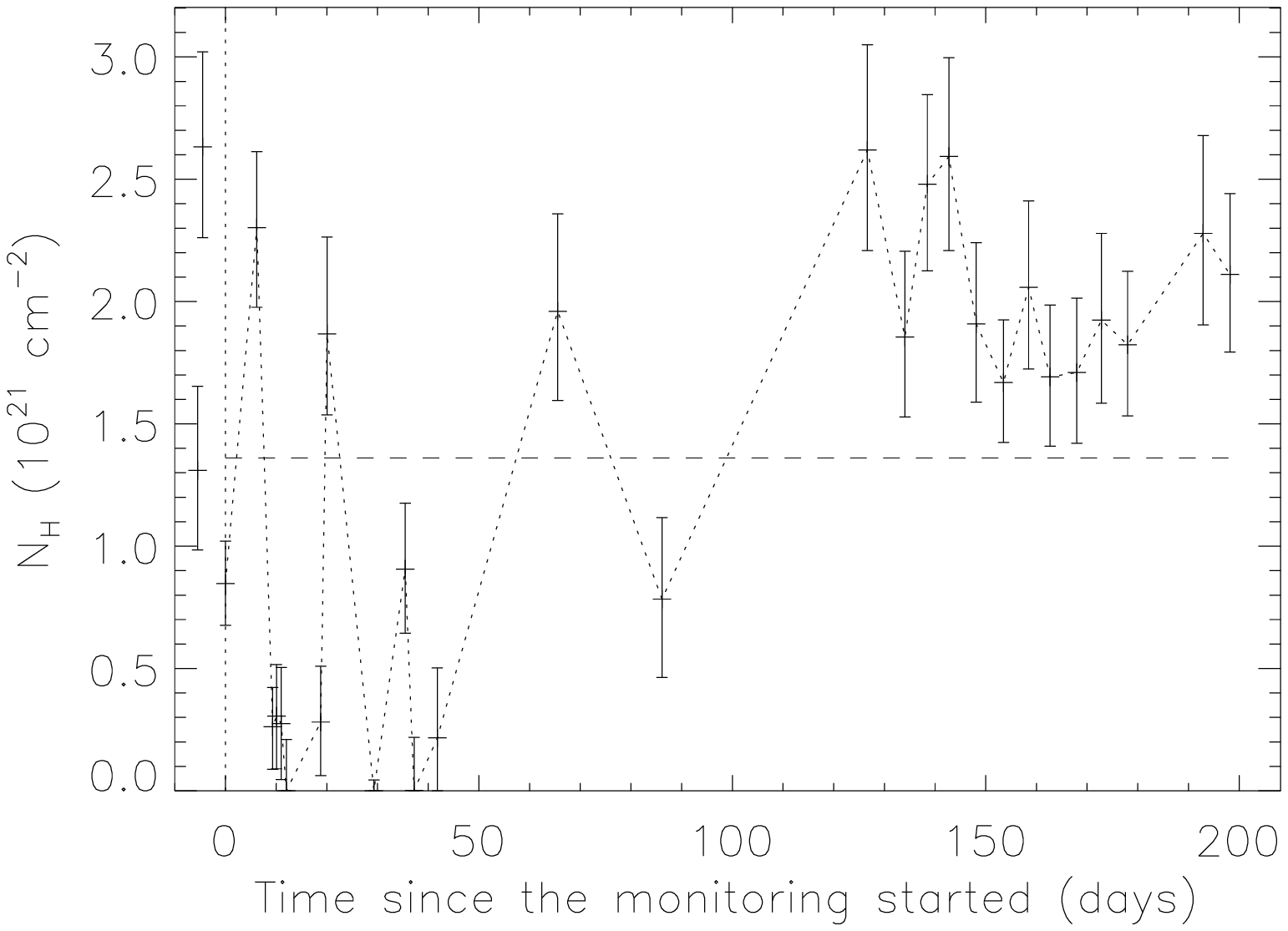}
\includegraphics[width=2.8in, height=2.0in]{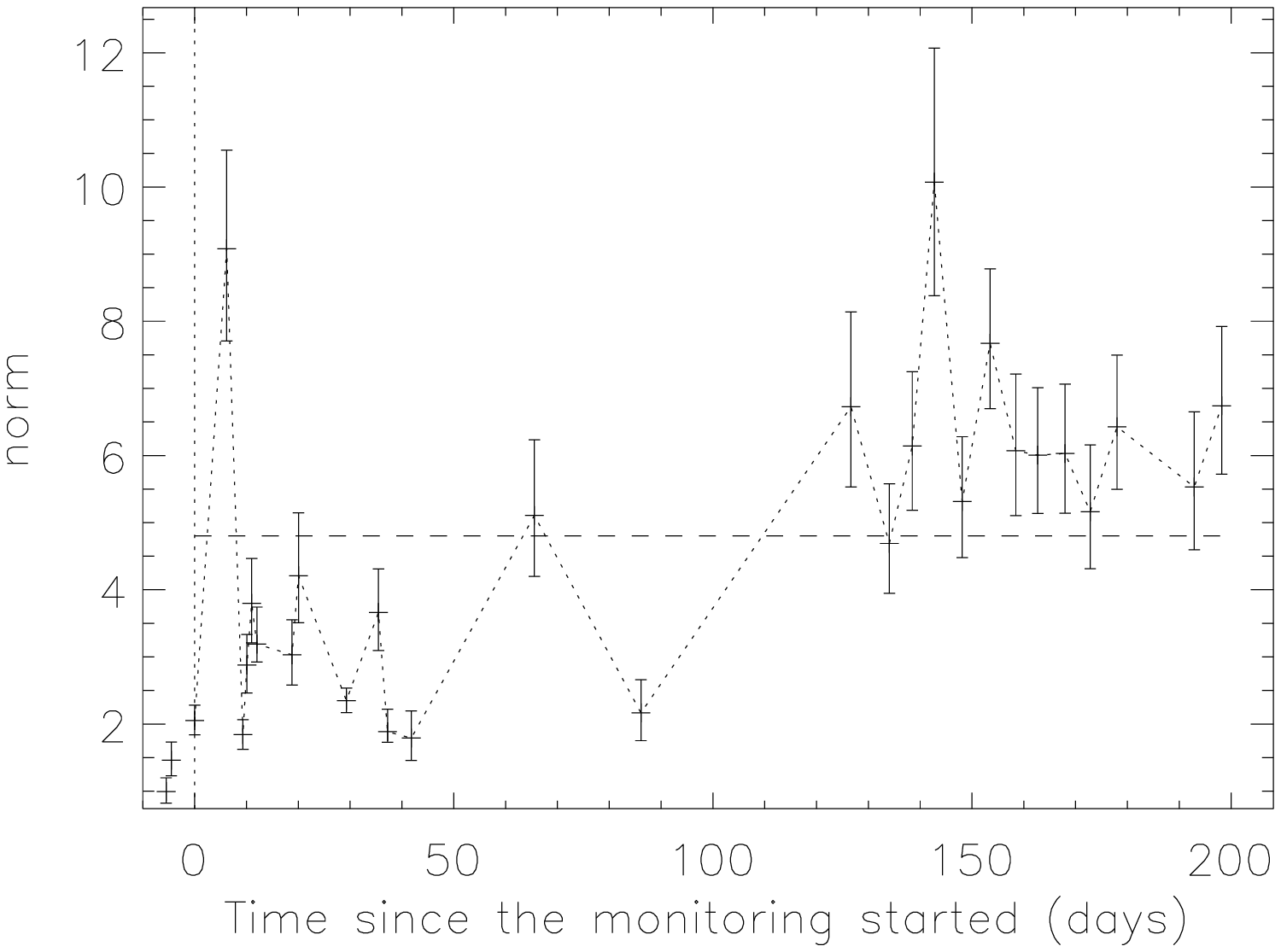}
\includegraphics[width=2.8in, height=2.0in]{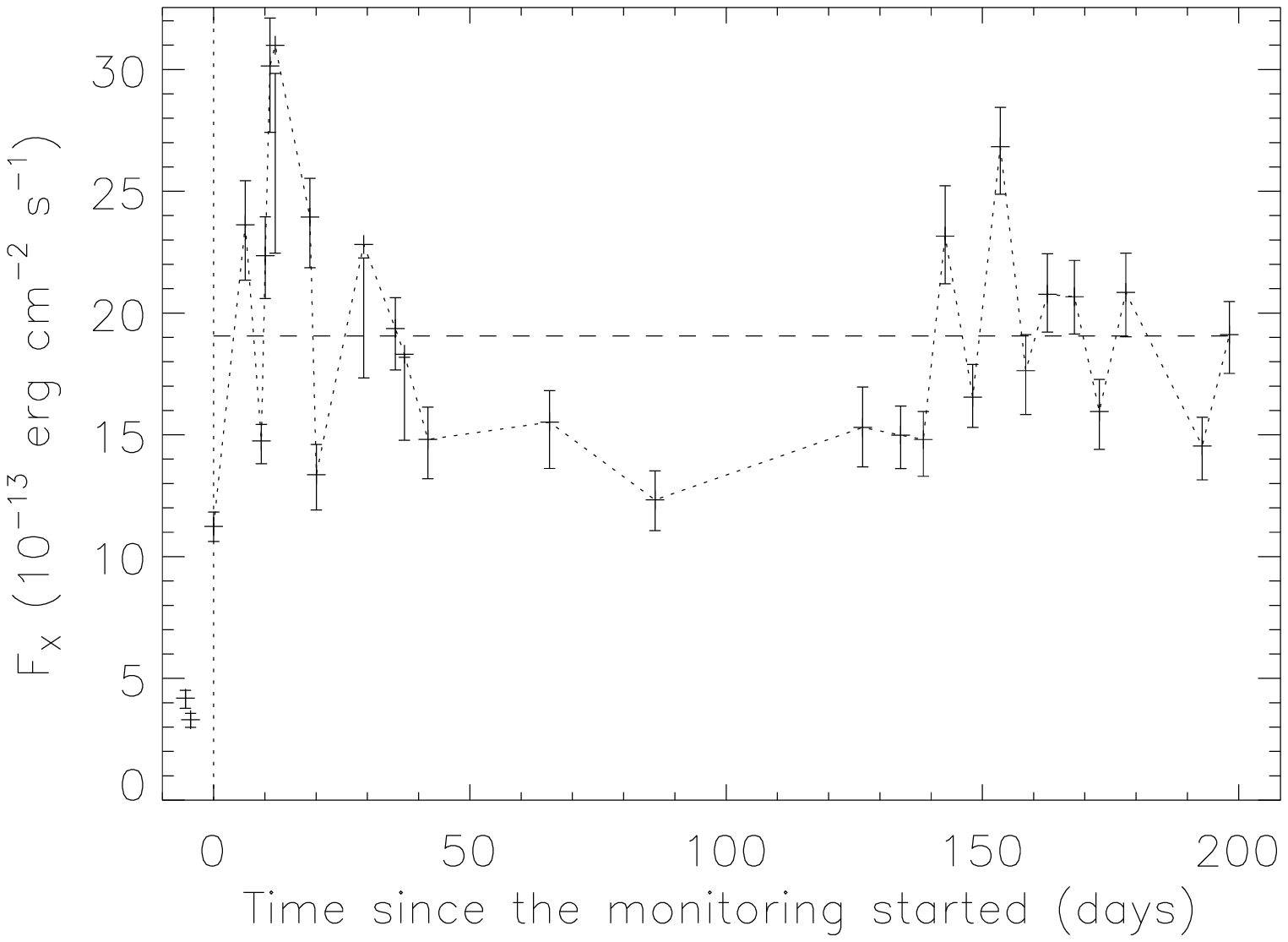}
\end{center}
\caption{Results from the fits to the \Swift spectra of \AG with
the CSW model for the data sets with higher than 100 X-ray counts.
Shown are the time evolution of
the X-ray absorption (N$_H$), model normalization (norm) and
observed X-ray flux (0.3 - 3 keV). The horizontal dashed line
indicates the mean value for the corresponding parameter.
The results for both observations in 2013 are given
at some fiducial negative time.
}
\label{fig:swift_csw}
\end{figure}

\SwiftE. 
It is then interesting to see whether the CSW model with nominal wind
and binary parameters (see above) can match the
X-ray emission from \AG as observed with \Swift in 2013 and 2015. To
do so, we adopted the same two-step approach as in the case of
discrete-temperature models (Section~\ref{sec:1T_model}).

First, we fitted simultaneously those spectra that have more than 100 
X-ray counts  allowing the O, Ne, Mg, Si and Fe abundances to vary.
The derived abundance values (with $1\sigma$ errors) are:
O $= 1.34^{+0.09}_{-0.12} $,
Ne $= 0.00^{+0.02}_{-0.00} $,
Mg $= 0.11^{+0.14}_{-0.11} $,
Si $= 0.55^{+0.18}_{-0.17} $,
Fe $= 0.55^{+0.05}_{-0.05} $
~(the values are with respect to the solar abundances; 
\citealt{ag_89}).
Second, each individual spectrum was fitted with the same CSW 
model having abundance values fixed to those derived in the first 
step. The free parameters for the individual fits were the amount of 
X-ray absorption and normalization parameter.

As in the case with discrete-temperature models 
(Section~\ref{sec:1T_model}), the quality of the fits to individual 
spectra was very good and some fit results are shown in 
Figs.~\ref{fig:swift_csw} and ~\ref{fig:spec_swift}.
It is worth noting that in the framework of the CSW model these
results suggest that the amount of shocked plasma in 2013 was about
the same as in 1993 (\Rosat observation). However, it has increased 
considerably in 2015 and not only that but it varies on relatively
short time scale (days). This follows from the values of the
normalization parameter ($norm$) which is proportional to the emission
measure of the hot plasma. Such changes require changes in the
mass-loss rates at least by a factor of 2 ($norm \propto \dot{M}^2$;
see above and also \citealt{zh_07} for details). As in the case 
of discrete-temperature models, the amount of X-ray absorption varies
considerably with time. 
 
Following the approach adopted in the discrete-temperature model
fitting, we considered the case of CSW model with common (being the
same) X-ray absorption for all the data set. This is equivalent to
having no variability of the X-ray absorption with time. We note that
the quality of the fits in this case was acceptable as well although
lower than in the standard CSW model case discussed above. Some
fit results are given in Table~\ref{tab:swift} which also illustrate
the considerable variability of the amount of shocked plasma needed to
explain the observed variability in the X-ray emission of \AGE.
We will return to these results in Section~\ref{sec:xray_origin}.

\begin{figure*}
\begin{center}
\includegraphics[width=2.8in, height=2.0in]{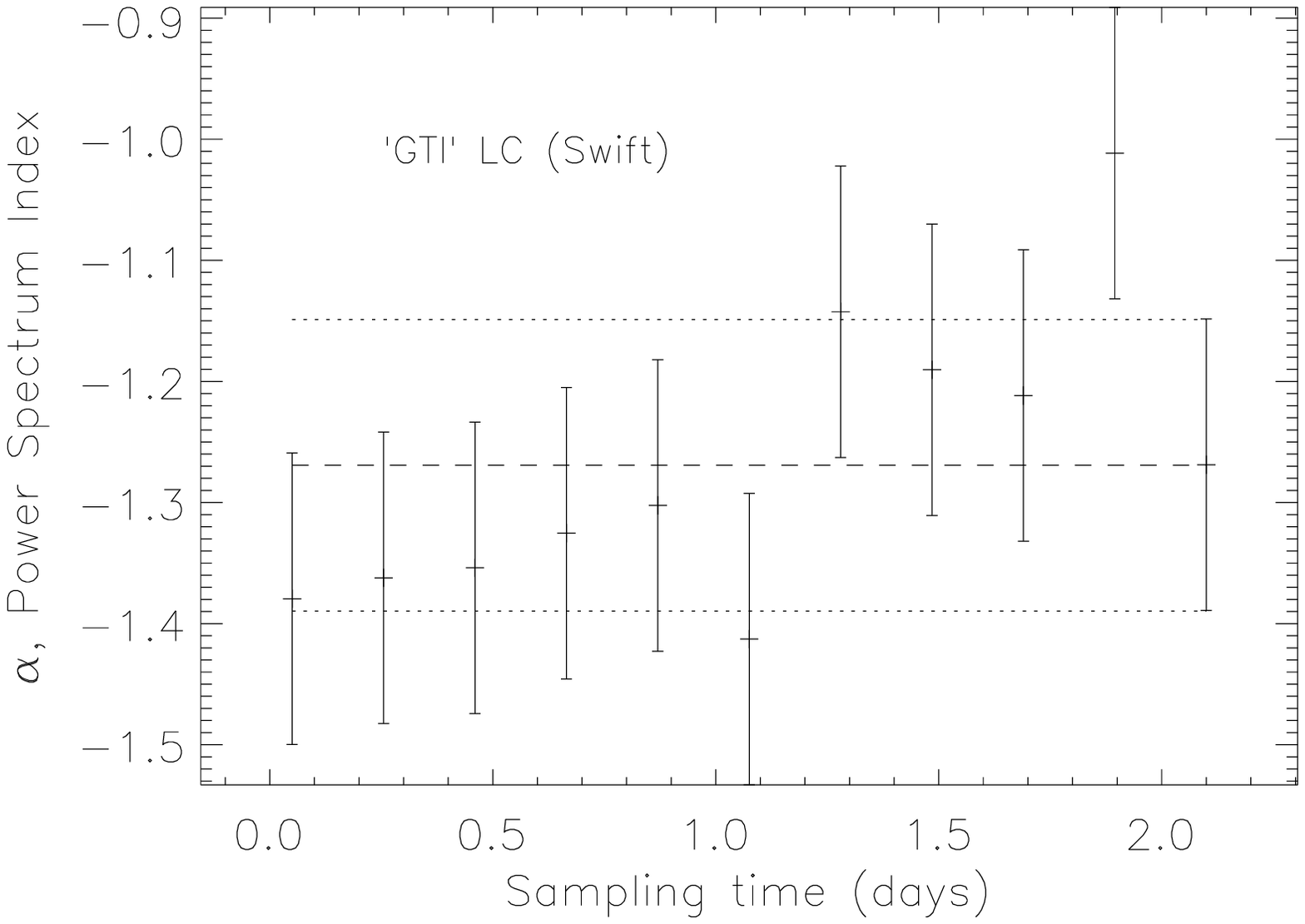}
\includegraphics[width=2.8in, height=2.0in]{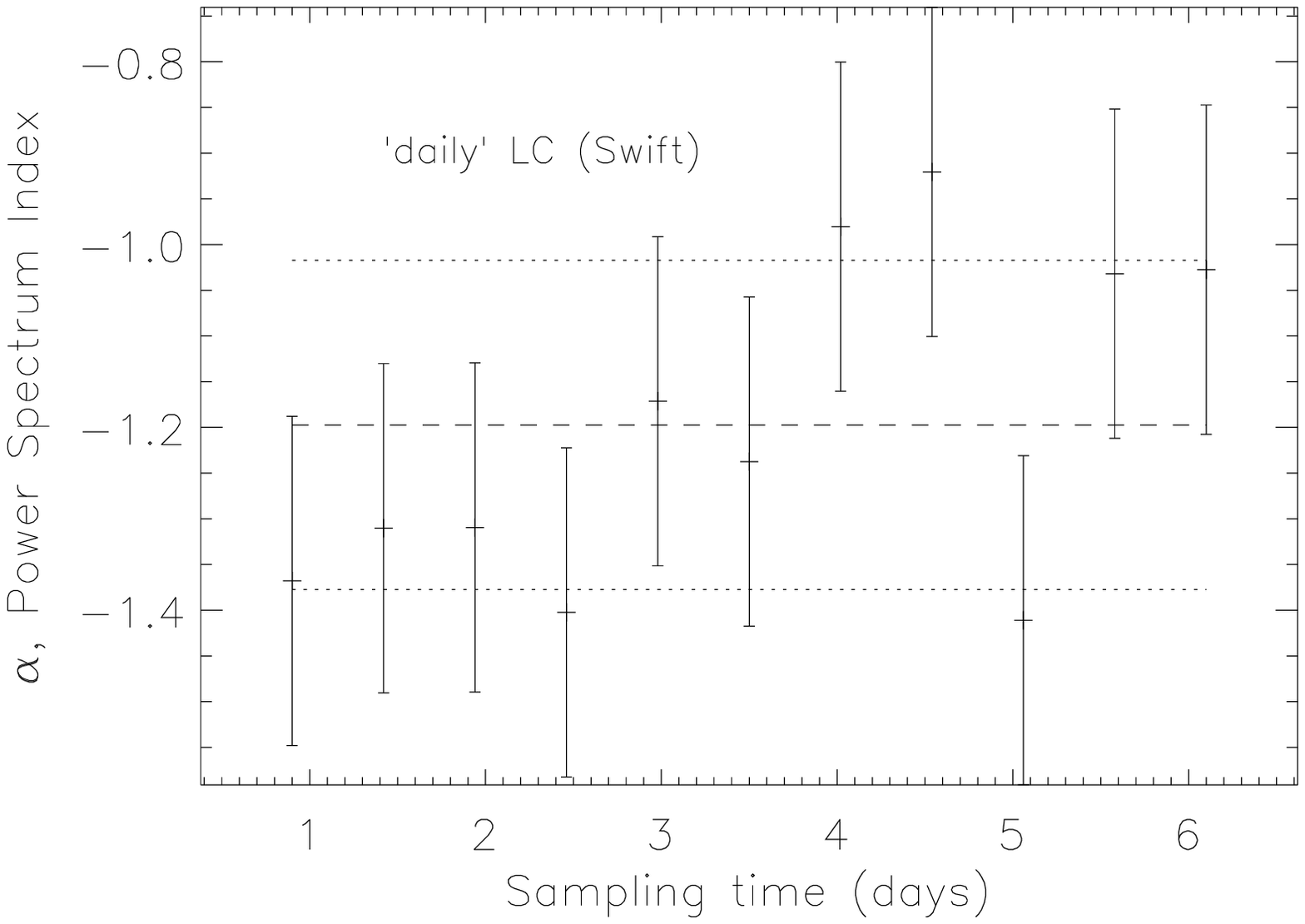}
\includegraphics[width=2.8in, height=2.0in]{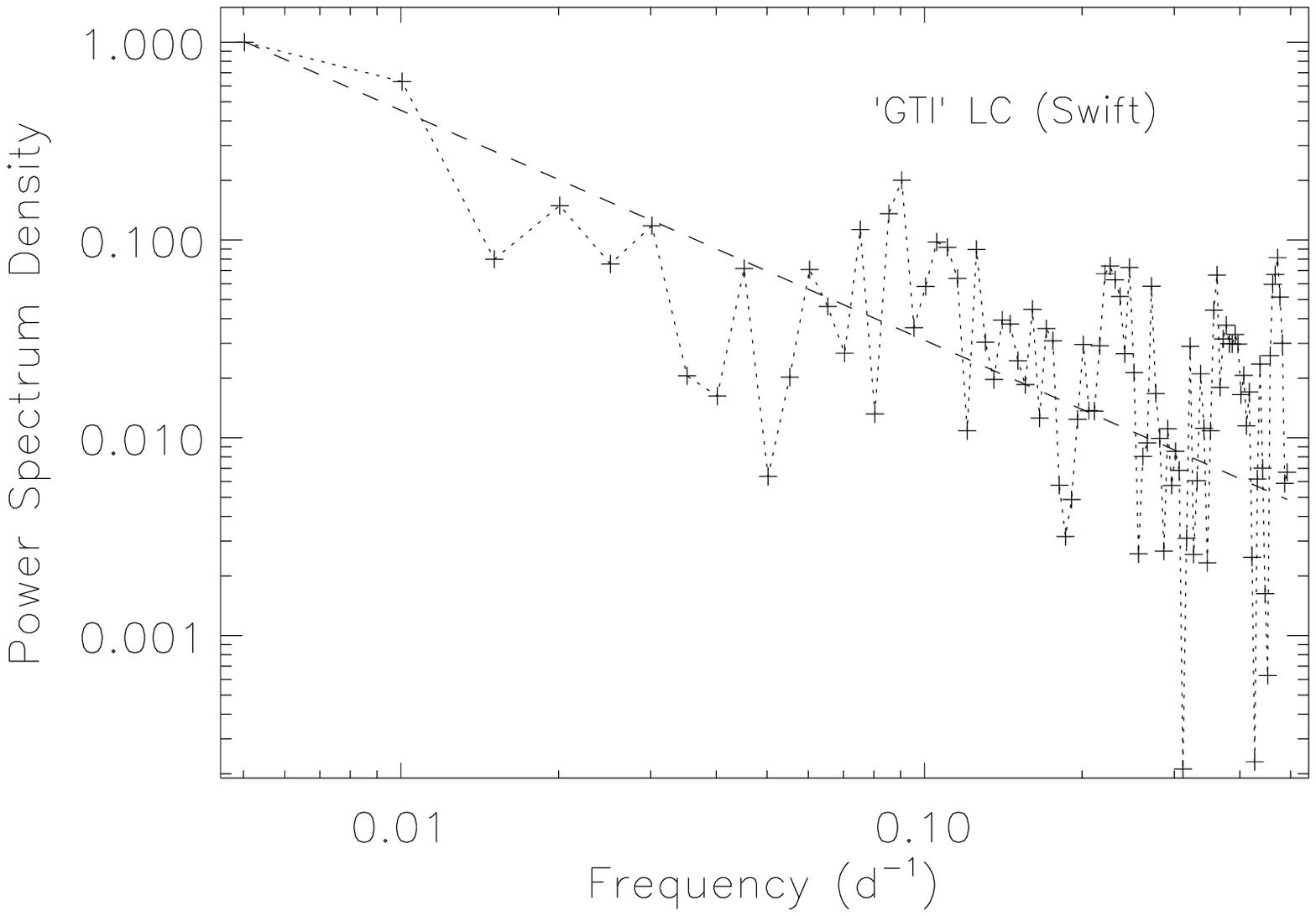}
\includegraphics[width=2.8in, height=2.0in]{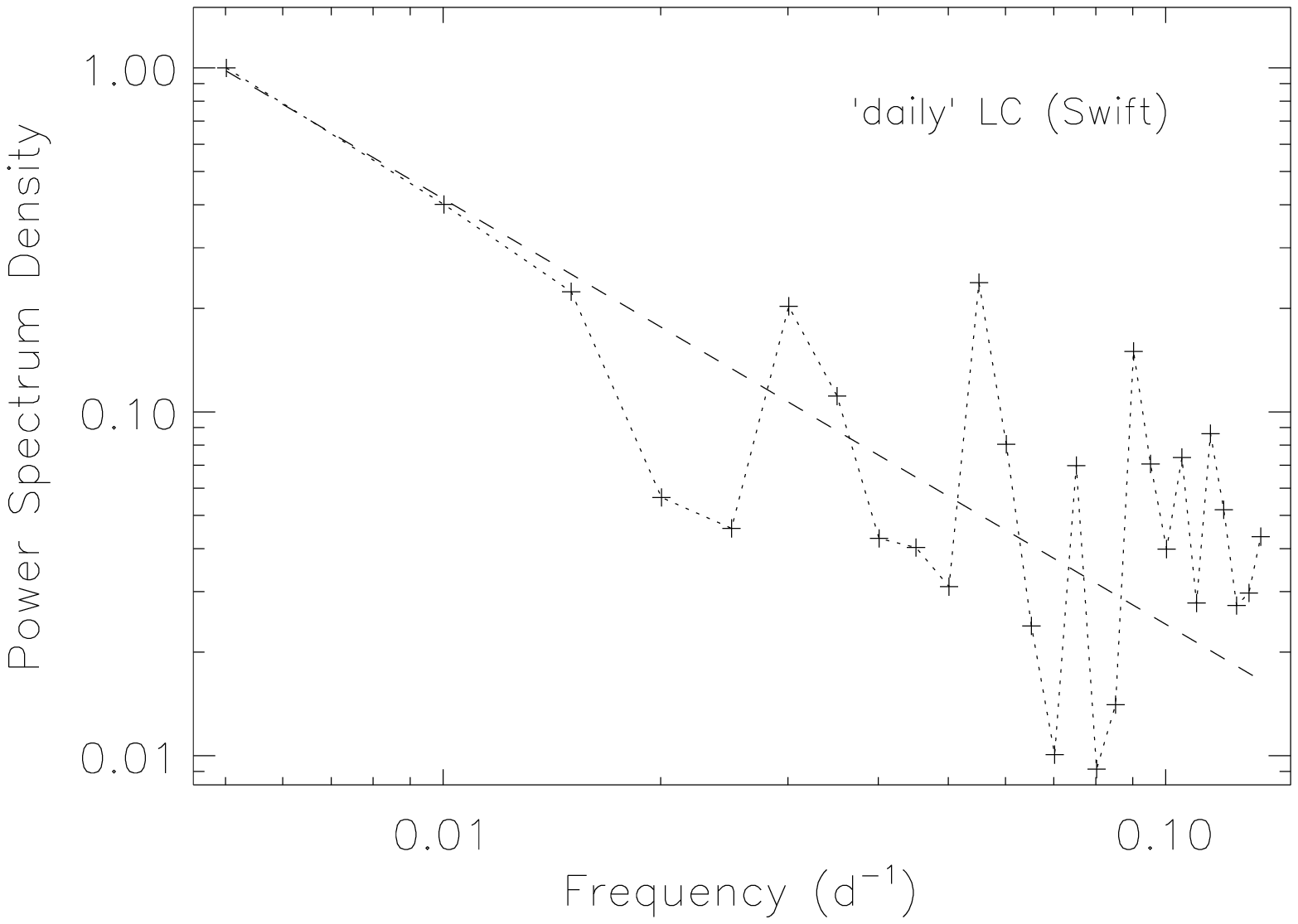}
\end{center}
\caption{
{\it Upper panels.}
The power-law index derived from the fits with a power-law
function to the power spectrum (PS) of the X-ray LCs (PS $\propto
f^{\alpha}$, $f = 1/\mbox{time}$). The mean value for the results
with different sampling time of the LC is shown by a dashed line. The
dotted lines mark its $1\sigma$ (standard deviation) confidence
interval.
{\it Lower panels.}
Examples of the model fits (dashed line) to the power spectrum 
(in log scale) for the cases with the respective  mean value of 
the sampling-time ranges 
(1 day for the `GTI' LC, $\alpha = -1.16\pm0.06$; 
3.5 days for the `daily' LC, $\alpha = -1.24\pm0.10$). 
The PS is normalized to its maximum value for presentation.
}
\label{fig:alpha}
\end{figure*}

\section{Origin of the X-ray emission}
\label{sec:xray_origin}
Our analysis of the \Swift data on \AG showed that its X-ray spectrum 
can be presented by emission from an optically thin plasma with a
temperature of a few 10$^6$~K (Section~\ref{sec:1T_model};
Fig.~\ref{fig:swift_1T}). This means that \AG is of the class $\beta$ 
of the X-ray sources amongst symbiotic stars as proposed by 
\citet{murset_97}. But, probably the most important result from the 
\Swift observations is the considerable X-ray variability that is 
present on time scales of days (and in some cases even hours;
Section~\ref{sec:xray_lc}; Fig.~\ref{fig:totalLC}; 
Appendix~\ref{append:X-ray}).

We recall that \citet{murset_95} suggested that the X-ray
emission from \AG is related to colliding stellar winds in binary
system. In this study, we showed that the \Rosat spectrum (1993 June) 
of \AG can well be presented by X-ray emission from CSWs in binary 
system. The adopted values of the stellar wind and orbital parameters 
in the CSW model correspond to those deduced from analysis in other 
spectral domains (e.g., \citealt{schmutz_96}; \citealt{fekel_00}).

On the other hand, the X-ray emission from \AG in 2013 and 2015 
(\SwiftE) does not seem to originate from CSWs. The most solid 
argument against the validity of the CSW picture in this case is the 
considerable X-ray variability that is present on time scales 
smaller than the
CSW dynamical time in AG Peg ($t_{dyn} = a/V = 3.7$~days;
$a$ is the binary separation  and $V$ is the wind velocity), 
i.e. on time scales of a couple of days (and in some cases even hours). 
This requires 
appreciable changes of the mass-loss rates by 20-50\% and even
by a factor of 2 on the same time scale.
Similarly, considerable changes  of the X-ray absorption are deduced
from analysis of the \Swift spectra in 2015 as indicated by 
Figure~\ref{fig:swift_csw} and Table~\ref{tab:swift}. The difference 
in X-ray absorption on the {\it point-by-point} basis is at the 
$(2-4)\sigma$ level with respect to zero (`identical' N$_H$ values), 
especially, in the first 40 days (July to mid-August 2015;
the average time interval between the data points is 1.1 days) 
after the 
\Swift monitoring has started.
We note that such X-ray absorption 
changes are hard to expect in the framework of the CSW picture. 
It is so since the CSW region is a 3D object with a size
at least as large as the size ot the binary orbit ($> 2$~au in \AGE).
So, to `hide' such a large structure and/or dissolve the 'curtains' 
(increase and/or decrease the X-ray absorption) on time scales of days 
does not seem realistic. Namely, the physical agent that causes such
changes on a dynamical time of a day should have a bulk velocity
higher than 3,000\kms ($> 2$ au / 86,400 s). So, such fast changes 
could indicate a size of the X-ray emission region {\it much smaller} 
than that of the CSW region in \AGE. Another characteristic of the 
X-ray light curves of \AG might be also pointing in that direction.

Namely, we tried to estimate the power spectrum of the X-ray LCs
obtained in the \Swift observations of \AGE. 

We recall that the data points in the \Swift LCs (`daily' LC and `GTI'
LC) are not equidistant
in time (see Section~\ref{sec:xray_lc}). For the power spectrum of a 
LC, we thus used a linear interpolation to `project' the LC onto a 
grid with equal time step. Then, we made use of the FFT (Fast Fourier 
Transform) algorithm to calculate the power spectrum (PS). Finally, 
we fitted a power-law function to the resultant power spectrum of the 
LC under consideration: PS $\propto f^{\alpha}$, $f =
1/\mbox{time}$. Interestingly, the fitted power-law index had negative 
values: $\alpha \approx -1$. To check how this result depends on the
sampling time of the LC, we explored a range for the 
time bins of the resultant `equidistant' LC. This range was chosen
trying to avoid `extreme' oversampling of the original LC as well as
to avoid having just few data points in the resultant one.
Namely, the range covers about 70-75\% of the number of time steps 
(intervals) in the original LCs as only the 10-15\% of the shortest 
and of the longest time steps were excluded.
The corresponding fit results are (mean value followed by the standard 
deviation in parentheses; Fig.~\ref{fig:alpha}): 
$\alpha = -1.20\  (0.18)$ for the `daily' LC; 
$\alpha = -1.27\  (0.12)$ for the `GTI' LC.

In order to test this result, we adopted the following approach. We
simulated a LC that has a power spectrum with the same $\alpha$-value 
as derived for the observed LC. A normal (Gaussian) noise was added
too.  The simulated LC covers a time 
interval from a few hundreds to more than a thousand days. Using the
time grid of the observed LC and a randomly chosen LC start time, we 
extracted a large number (e.g., a thousand) of `theoretical' LCs. Each
of these were processed exactly in the same manner as the observed LC.
The result was that the derived values of $\alpha$ were about normally 
distributed with mean value very close to that of the observed LC,
in fact, within one standard deviation. This numerical test gave us
additional confidence in the derived $\alpha$-values of the power 
spectrum of the X-ray LCs of \AGE.

It is interesting to note that the values of $\alpha$ derived from the
X-ray variability of \AG are close to $\alpha = -1$ which is
characteristic of the so called flicker noise (or flickering) found in
many astronomical objects and typical for variety of physical
phenomena (e.g., see \citealt{press_78}). In astronomical objects,
flickering is usually found in systems that harbour degenerate objects
like black holes, neutron stars, white dwarfs and is related to
accretion processes. Then, the power spectrum of the LCs of \AG as
observed with \Swift in 2015 might be also indicating that the X-rays 
in this symbiotic system do not originate in CSWs.

Thus, 
to explain the observed X-ray properties of \AG in 1993-2013-2015, we
could assume that (a) CSWs did play a key role in generating the X-ray
emission in 1993 (\RosatE); (b) in the period before 2013, \AG
switched to a different mode, that is, the stellar wind of the hot
component (a white dwarf; WD) has seized and the accretion of the cool
star wind onto the WD started. 
We recall that such a change in \AG has been suggested earlier 
(see \citealt{zamanov_95} for details). However, if this were indeed 
the case, some `drastic' changes in the optical spectrum of \AG should 
have occurred as well. Among others, there must be no signs of 
{\it strong} stellar wind from the hot component in this symbiotic 
binary. Interestingly, this is exactly what is found by 
{\color{blue}
Tomov, Stoyanov \& Zamanov  (2016, MNRAS, submitted) 
}
from the analysis of optical emission of \AG
as observed in 2015.
These authors suggest 
that a new phase in the evolution of \AG is under way, which finds 
further support from the analysis of the \Swift data on this object 
presented here.

However, it is important to note that the \Swift spectra of
\AG do not manifest signs of highly absorbed and strong emission
at high energies (above 2.4 keV). Such kind of emission is considered
typical for the so called $\delta$ and $\beta/\delta$ X-ray sources
among the symbiotic stars and is assumed to originate in the 
accretion-disk boundary layer \citep{luna_13}.

One reason for not detecting X-ray emission at high energies could be
purely technical: the really short exposure times of the \Swift 
spectra of \AGE. 

To explore this a bit more, we created a total
spectrum by summing up all the 71 spectra in 2015. 
The total spectrum shows that some weak emission is present in the
2 - 5 keV energy range. 
A thin-plasma model with at least two temperature components
(kT$_{low} \sim 0.5$ keV; kT$_{high}\sim 2.7$ keV)
could match that high-energy emission.
But, we emphasize it 
again that this exercise had purely technical meaning since it is not 
advisable (e.g., in physical sense the result might be misleading) 
to create a total spectrum for a source exhibiting high
variability with appreciable spectral changes as is the
case of \AG (see Fig.~\ref{fig:totalLC} and Table~\ref{tab:swift}).
The latter is the most likely explanation why the quality of the fit 
to the total spectrum using the two-temperature model was not high 
($\chi^2/dof = 150/96$).
Thus, {\it individual} X-ray spectra with good quality are needed to
reveal if high energy X-ray emission is present in this object and to
constrain its characteristics.

But as mentioned above, another reason for the lack of high energy 
X-ray emission could be 
related to the actual physical processes taking plays in \AGE.
 
We recall that thermal emission with relatively low plasma
temperatures (kT$\leq 1$ keV) is considered a sign of CSWs in the
symbiotic stars. But, an important ingredient is mandatory to have
this mechanism working. Namely, it is the presence of relatively
strong and fast stellar wind from the hot component,
which is established from observations in other spectral domains 
(UV, optical). Such was indeed
present around the time of the \Rosat spectrum of \AG
(\citealt{nuss_95}; \citealt{schmutz_96}). 
However, analysis by Tomov et al. (2016) of recent optical spectra of 
\AG showed that no such wind is present any more. 
 
But, let us consider the results from the optical (Tomov et al. 2016)
in a conservative way. Namely, had they {\it only} meant that the 
stellar wind of the hot component in \AG is now considerably less 
massive (i.e. harder to detect), the X-ray emission from the supposed 
CSW region in the 
system could have been substantially reduced (see the scaling law for 
CSWs discussed in Section~\ref{sec:csw_model}). Therefore, we would 
expect weaker X-ray emission in recent observations. Opposite to this, 
the X-ray emission from \AG has increased considerably compared to its 
level of about two decades ago. Thus even 
adopting a conservative approach, additional
component is needed to describe the observed X-ray properties of \AGE.
Moreover,
keeping in mind that the 
considerable X-ray variability of \AG is very 
hard to accommodate in the framework of a plausible ($=$ not too
speculative)  CSW picture and 
that its time characteristics resemble those of a flicker noise 
(flickering), it is our understanding that a different process is 
getting into play and accretion onto a degenerate hot companion seems 
the most likely to us.

Additional support for the probable importance of accretion processes
in \AG comes from its UV light curve. We note that analysis of the
complete set of UV observations of \AG with \Swift is beyond the scope
of this work: in detail analysis is postponed to another paper. 
Here,
we only mention the results for the data set with the longest exposure
(Obs 3; 2015 June 28; the observation identification is the same as
for the X-ray data; see Section~\ref{sec:data}). Namely, the UV light 
curve of Obs 3 clearly shows appreciable variability on time scales of
minutes and hours (Appendix~\ref{append:UV}; Fig.\ref{fig:uvLC}).
Such a UV variability is quite typical for the $\delta$ X-ray sources
among the symbiotic stars for which accretion is assumed to play 
an important role (see section 5.2 in \citealt{luna_13}).

Finally,
since \AG is likely in transition to a new evolutionary
phase (from a symbiotic nova to a classical symbiotic star;
Tomov et al. 2016), it might well be that it does not exactly
match any specific case in the classification scheme by
\citet{luna_13} of the X-ray sources among the symbiotic stars.
On the one hand, most of the X-ray emission of \AG is at energies
below 2 keV, thus, resembling a $\beta$ X-ray source. On the other 
hand, strong variability is present, thus, resembling sources fuelled 
by accretion processes. Also, the average X-ray luminosity of \AG
(L$_X > 10^{32}$ egrs s$^{-1}$; Table~\ref{tab:swift}) seems 
typical for a $\delta$  X-ray source (see section 5.2 in
\citealt{luna_13}), thus, again indicating likely importance of 
accretion processes.
We believe that further X-ray observations of \AG will be very 
helpful to follow the evolution of its properties and to finally
settle its classification status as X-ray source among the
symbiotic stars.

\section{Conclusions}
\label{sec:conclusions}
In this work, we analysed archive X-ray data (\Rosat and \SwiftE) on 
the symbiotic binary \AGE. The basic results and conclusions are as
follows.

(i) The X-ray emission of \AG as observed with \Swift in 2015 shows
considerable variability on time scale of days as variability on 
shorter time scales might be present as well. No periodicity is
present in the available data (2015 June - 2016 January). 

(ii) Analysis of the X-ray spectra obtained in 2013 and 2015 confirms
that \AG is an X-ray source of class $\beta$ of the X-ray sources
amongst the symbiotic stars. That is, its X-ray spectrum can be 
presented by emission from an optically thin plasma with a
temperature of a few 10$^6$~K.

(iii) 
The X-ray emission of \AG as observed with \Rosat (1993 June) 
might well originate
from colliding stellar winds in binary system. The 
CSW model with values that correspond to the stellar wind and orbital 
parameters deduced from analysis in other spectral domains (e.g.,
\citealt{schmutz_96}; \citealt{fekel_00}) can match very well the 
\Rosat spectrum of \AGE.

(iv) On the other hand, the characteristics of the X-ray emission of
\AG as observed with \Swift in 2013 and 2015 are hard to accommodate 
in the framework of the CSW picture. Namely, such are the considerable 
changes of the amount of X-ray emitting plasma and the X-ray 
absorption on time scale of one-two days. 

(v) Analysis of the \Swift light curves in 2015 shows that the power
spectrum of the X-ray variability in \AG can be matched by a power-law
function, PS $\propto f^{\alpha}$, $f =1/\mbox{time}$, with $\alpha
\approx -1.2$. This value is close to $\alpha = -1$, which is
characteristic of the flicker noise (or flickering) being typical for
accretion processes in astronomical objects. This, along with the
result mentioned above in (iv), is a sign that CSWs did not play a key
role for the X-ray emission from \AG as observed in 2013-2015 and a
different mechanism (probably accretion) is likely 
getting into play.

(vi) Further monitoring of the X-ray emission from \AG will be very
helpful to reveal {\it in detail} the physical picture in this 
fascinating symbiotic star. Also, uninterrupted X-ray observations 
with good quality will be very important to show whether the 
flickering is present on short time scales ($< 1$ day).

\section*{Acknowledgements}
This research has made use of data and/or software provided by the
High Energy Astrophysics Science Archive Research Center (HEASARC),
which is a service of the Astrophysics Science Division at NASA/GSFC
and the High Energy Astrophysics Division of the Smithsonian
Astrophysical Observatory. 
This research has made use of the NASA's Astrophysics Data System, and
the SIMBAD astronomical data base, operated by CDS at Strasbourg,
France.
The authors thank an anonymous referee for 
some comments and suggestions.

\bibliography{zht_agpeg.bbl}

\appendix

\section{Individual X-ray light curves of \AG}
\label{append:X-ray}
\begin{figure*}
\begin{center}
\includegraphics[width=2.24in, height=1.6in]{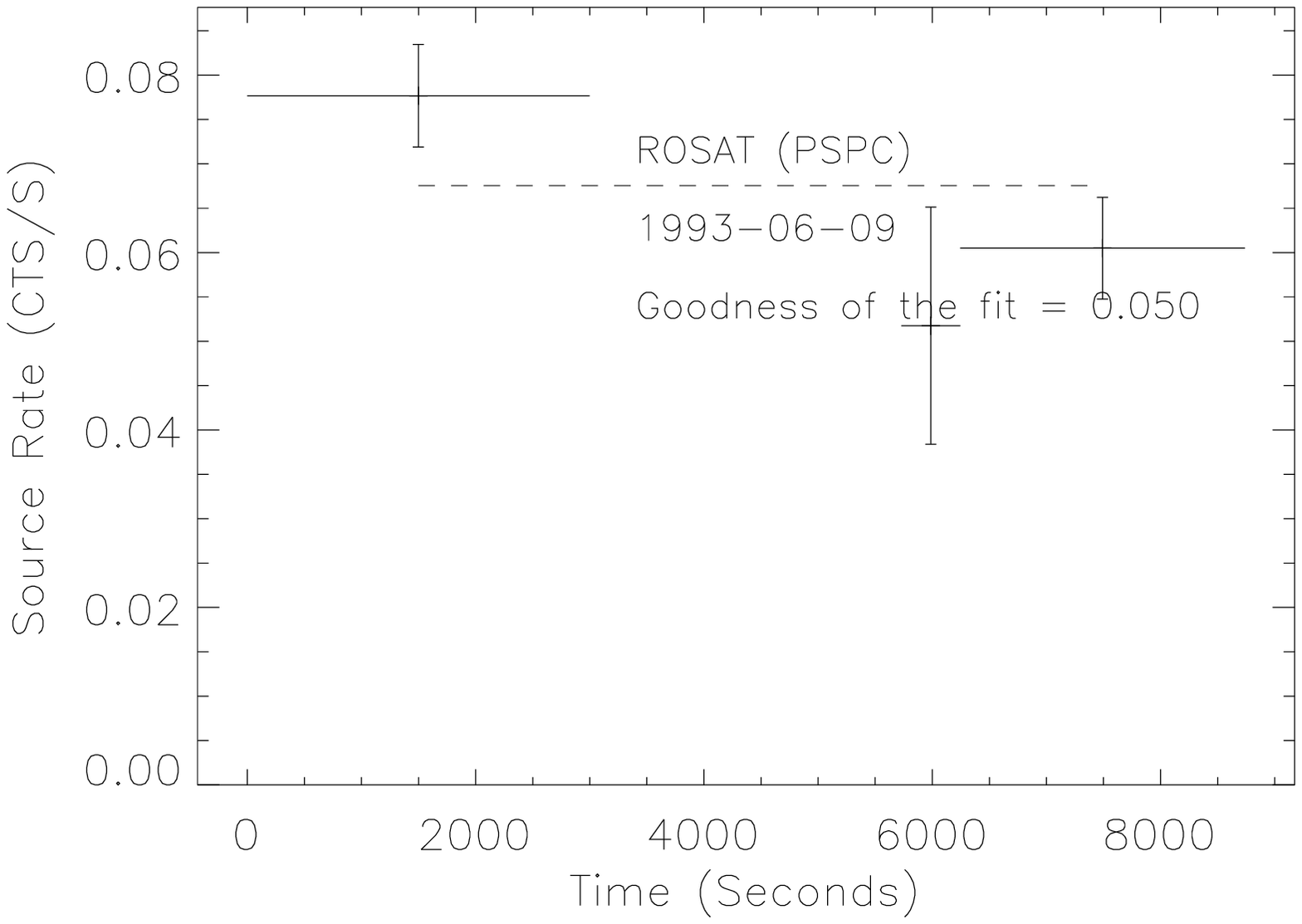}
\includegraphics[width=2.24in, height=1.6in]{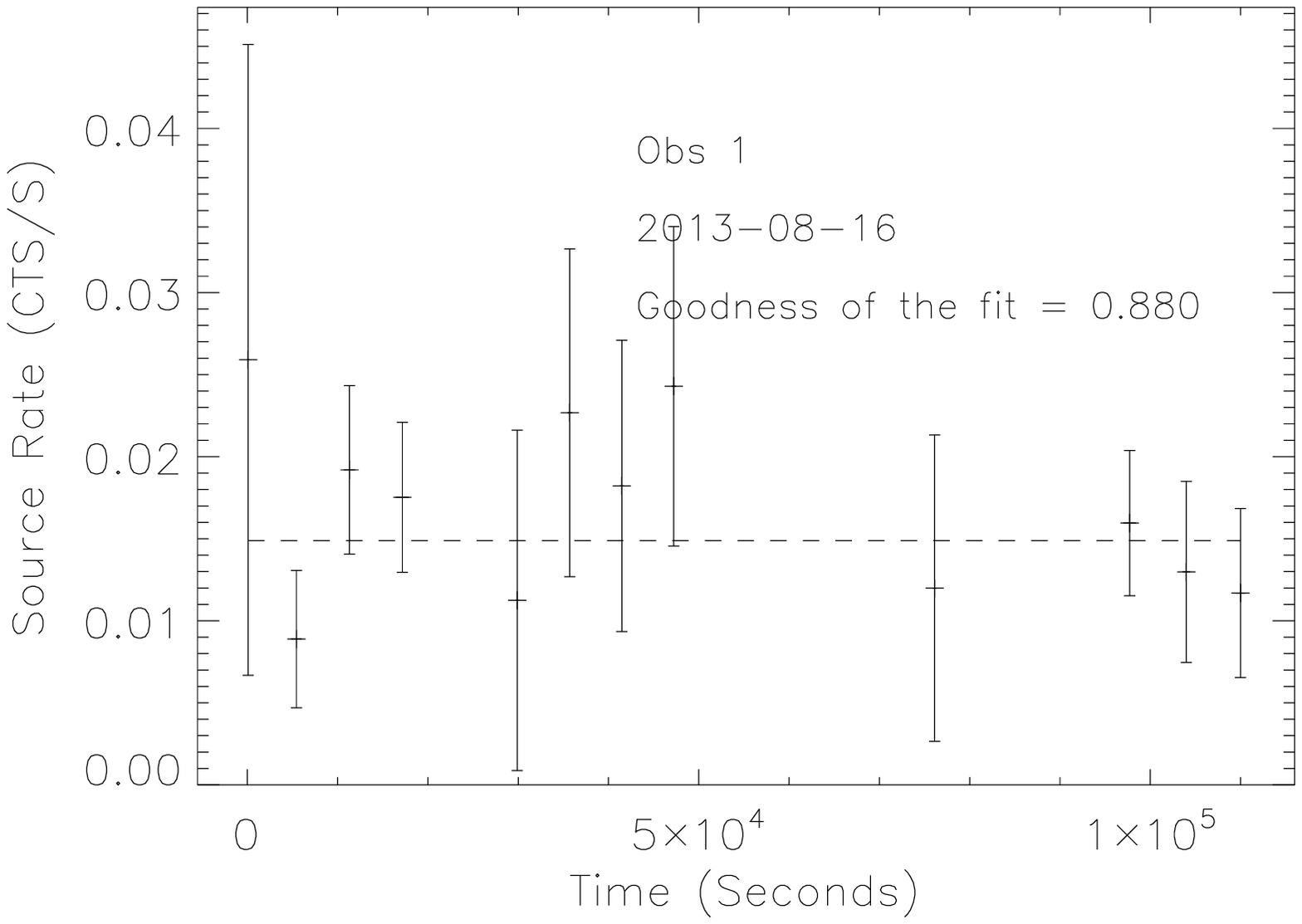}
\includegraphics[width=2.24in, height=1.6in]{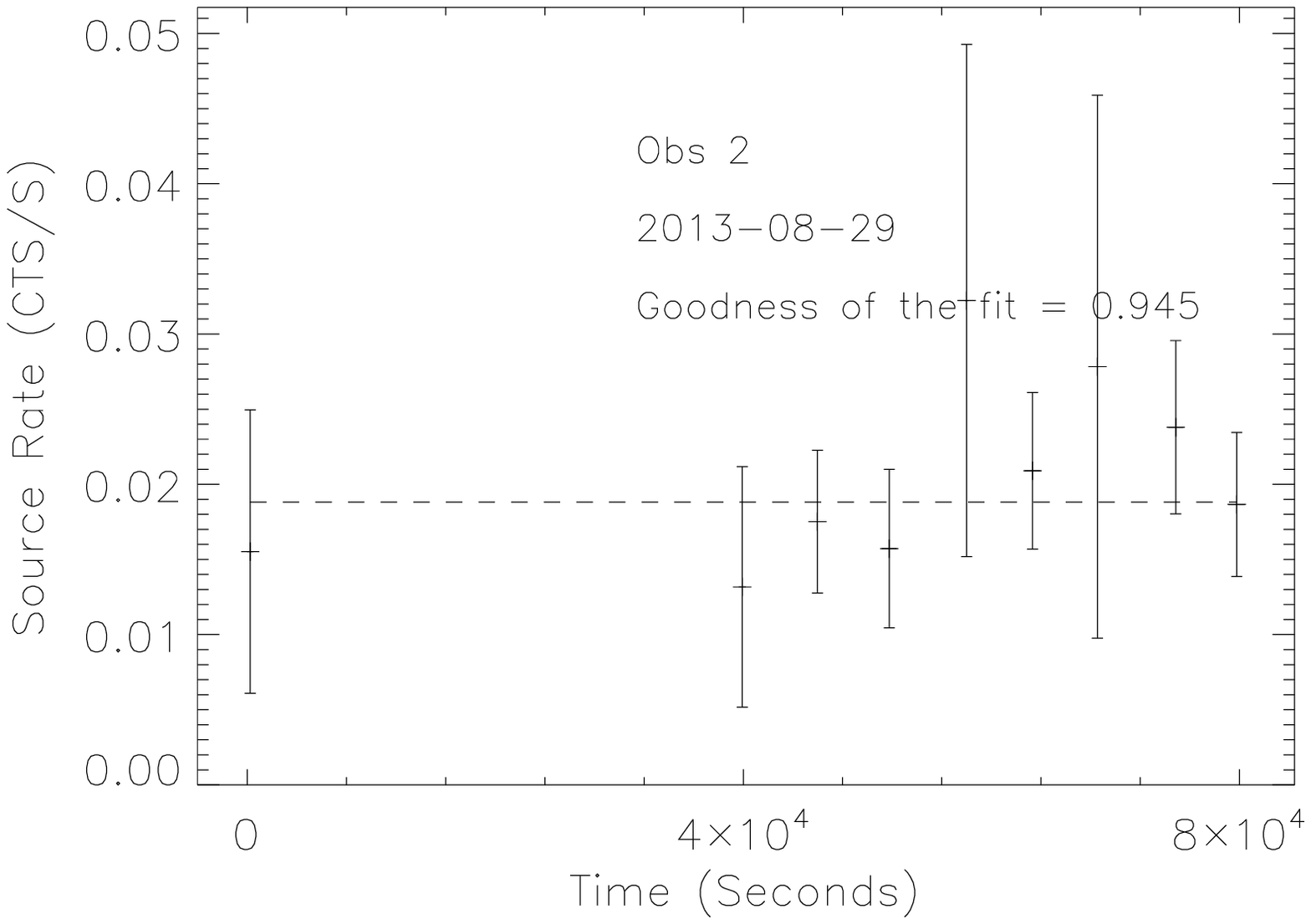}
\includegraphics[width=2.24in, height=1.6in]{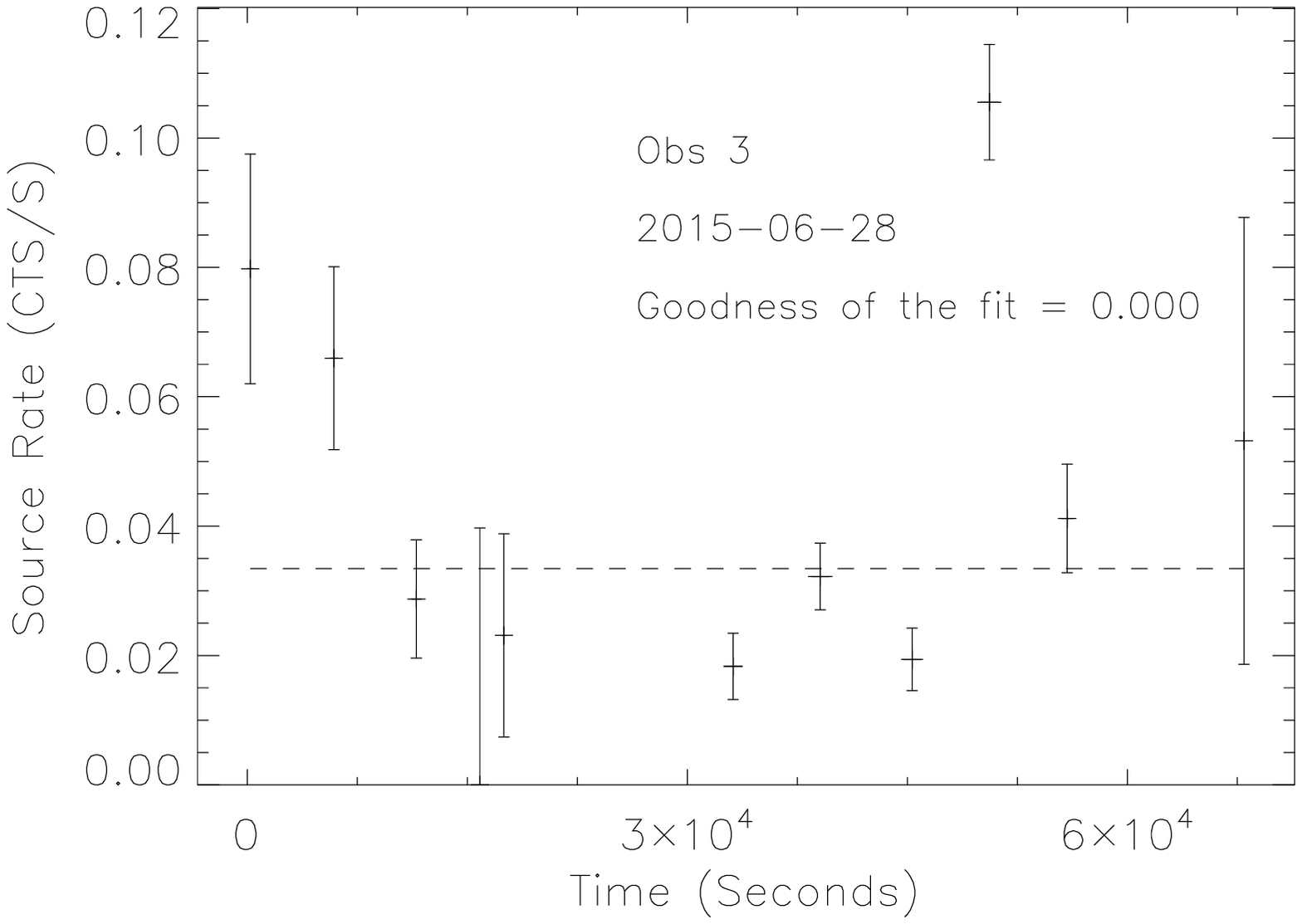}
\includegraphics[width=2.24in, height=1.6in]{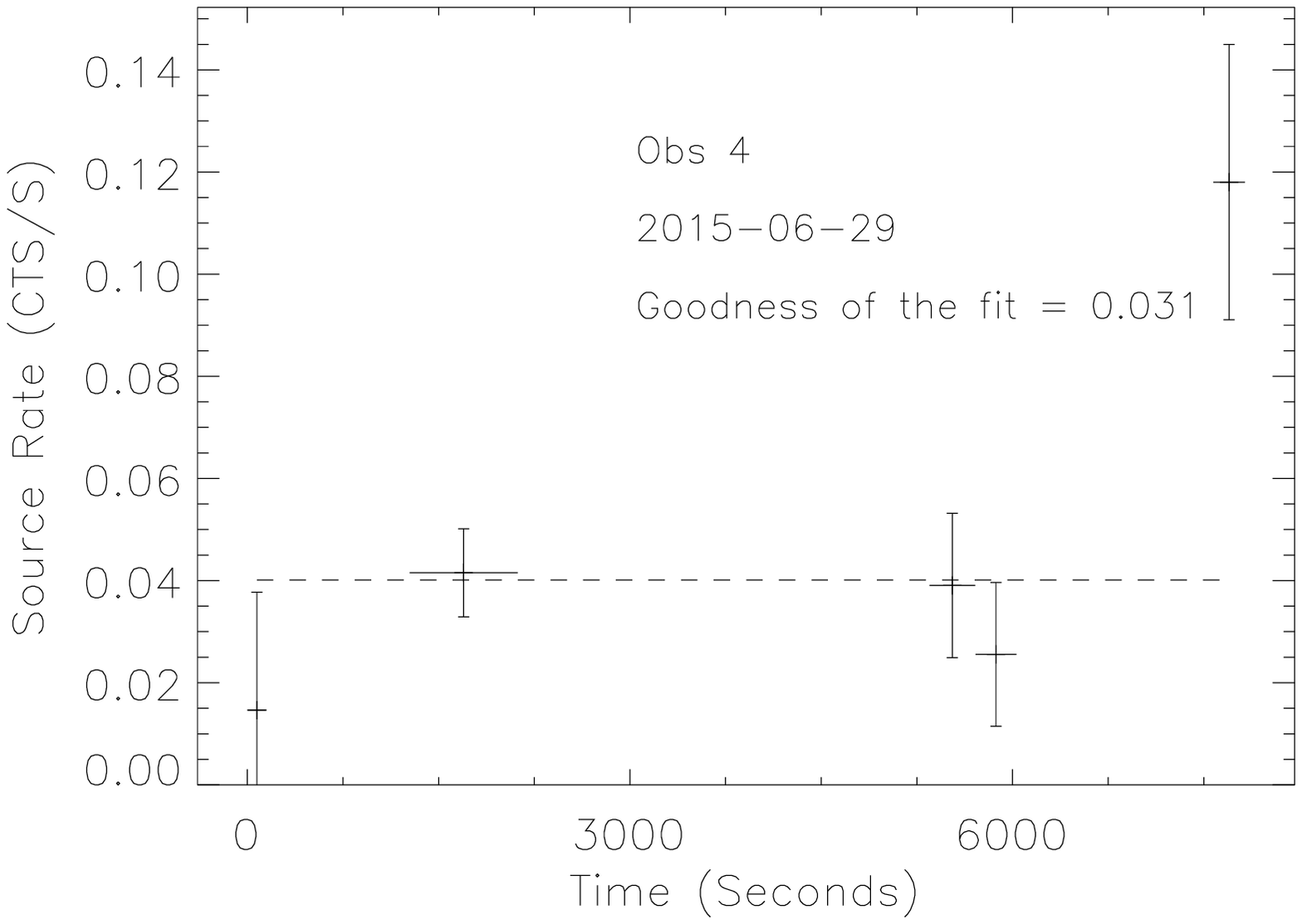}
\includegraphics[width=2.24in, height=1.6in]{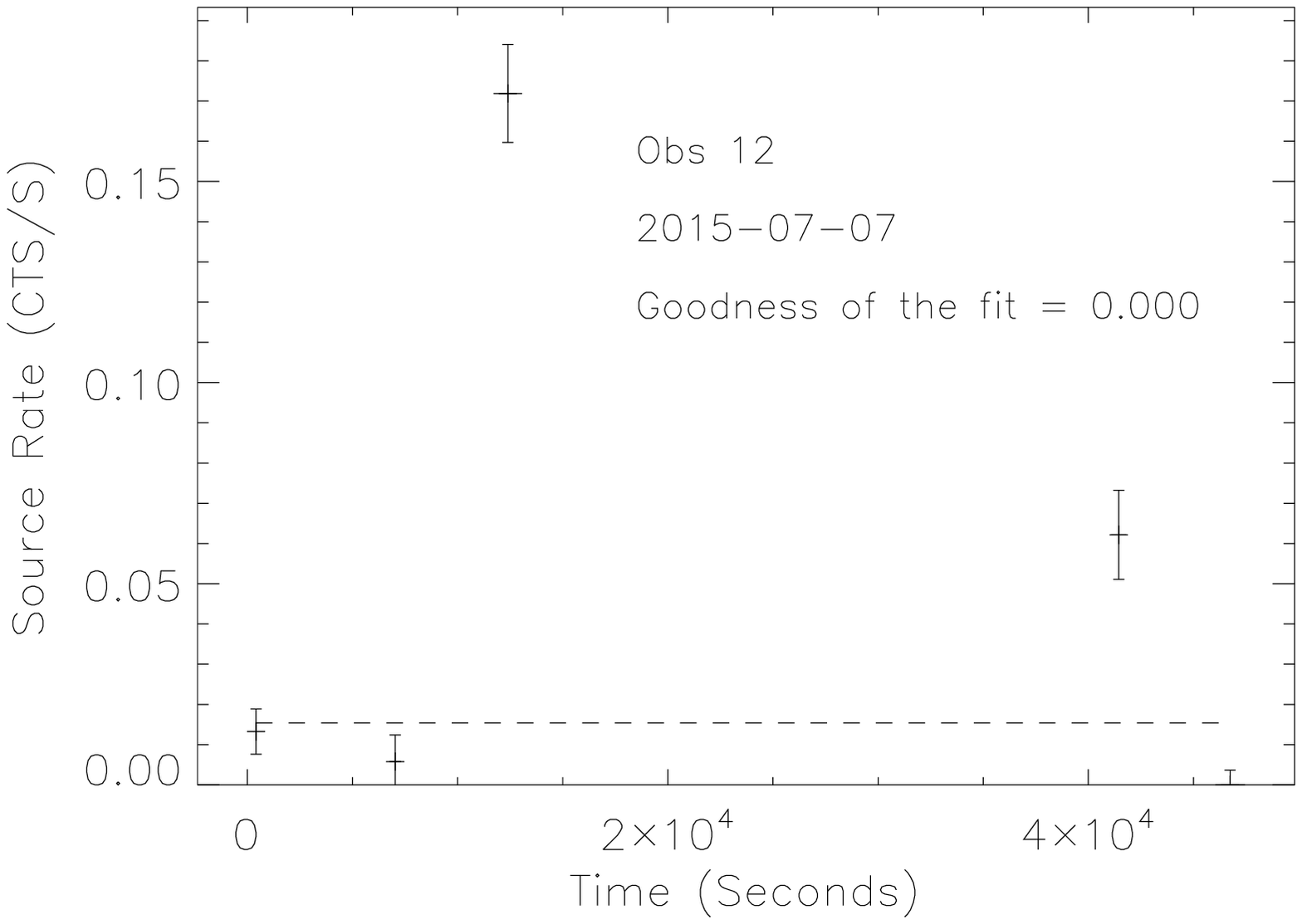}
\includegraphics[width=2.24in, height=1.6in]{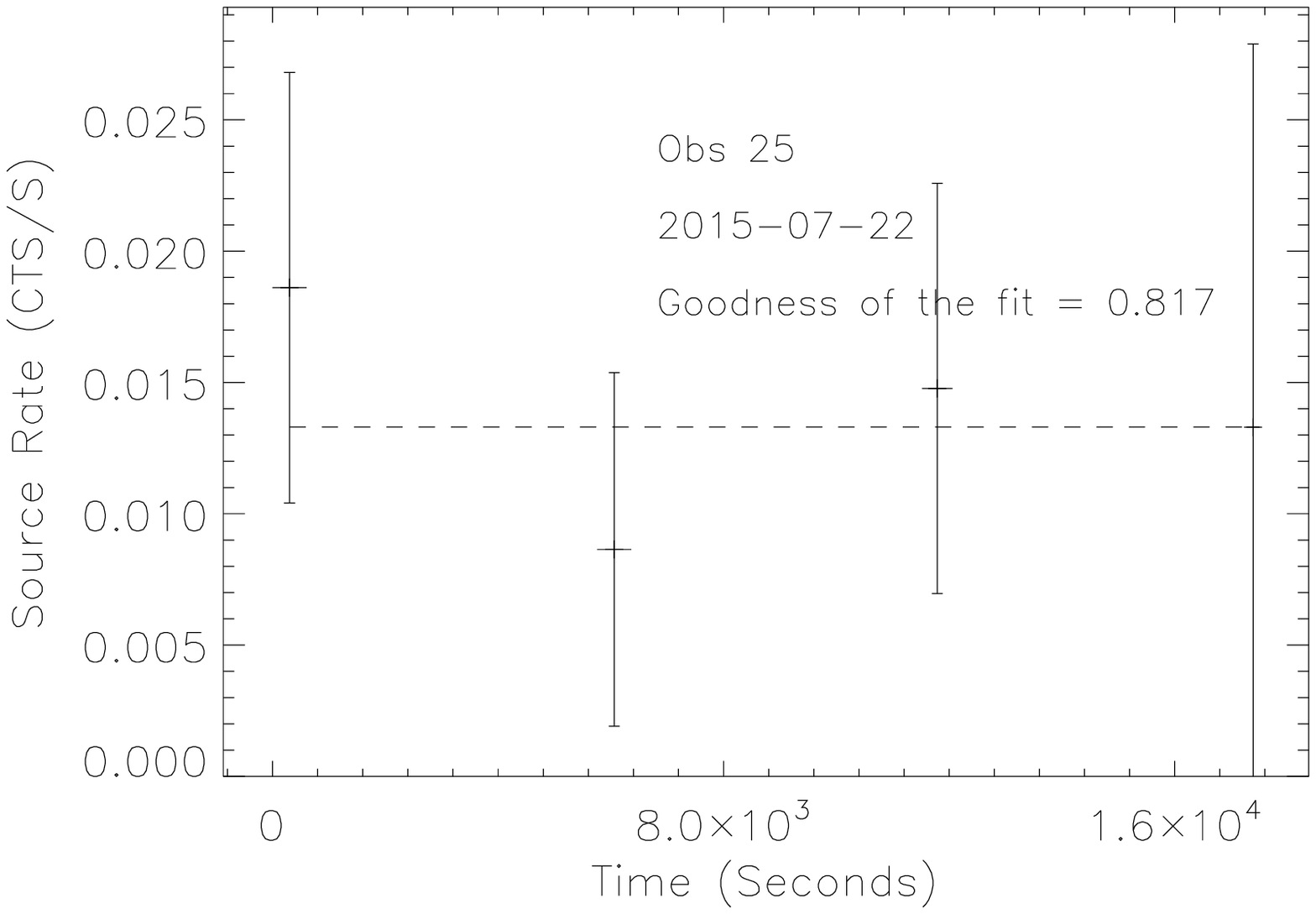}
\includegraphics[width=2.24in, height=1.6in]{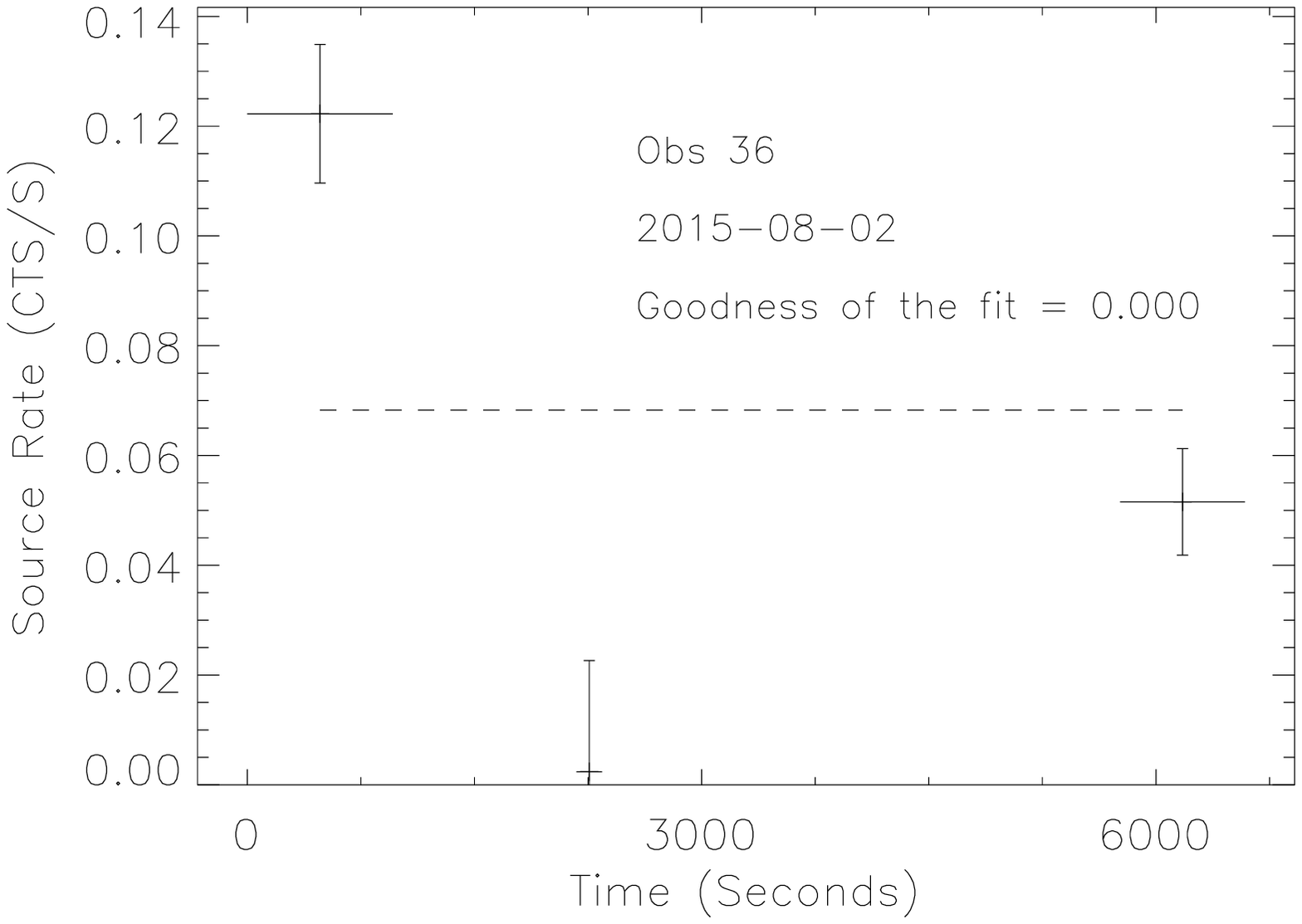}
\includegraphics[width=2.24in, height=1.6in]{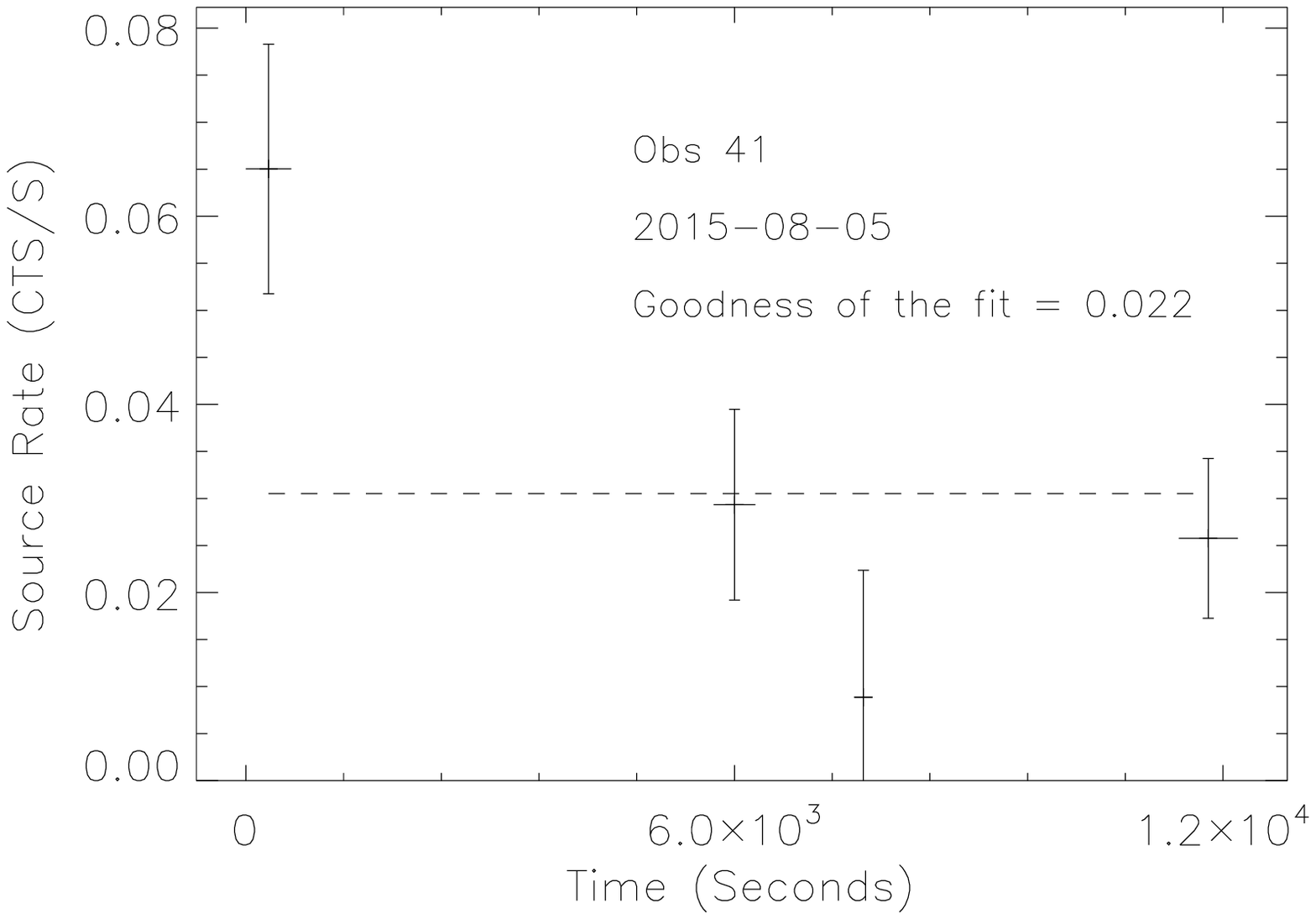}
\includegraphics[width=2.24in, height=1.6in]{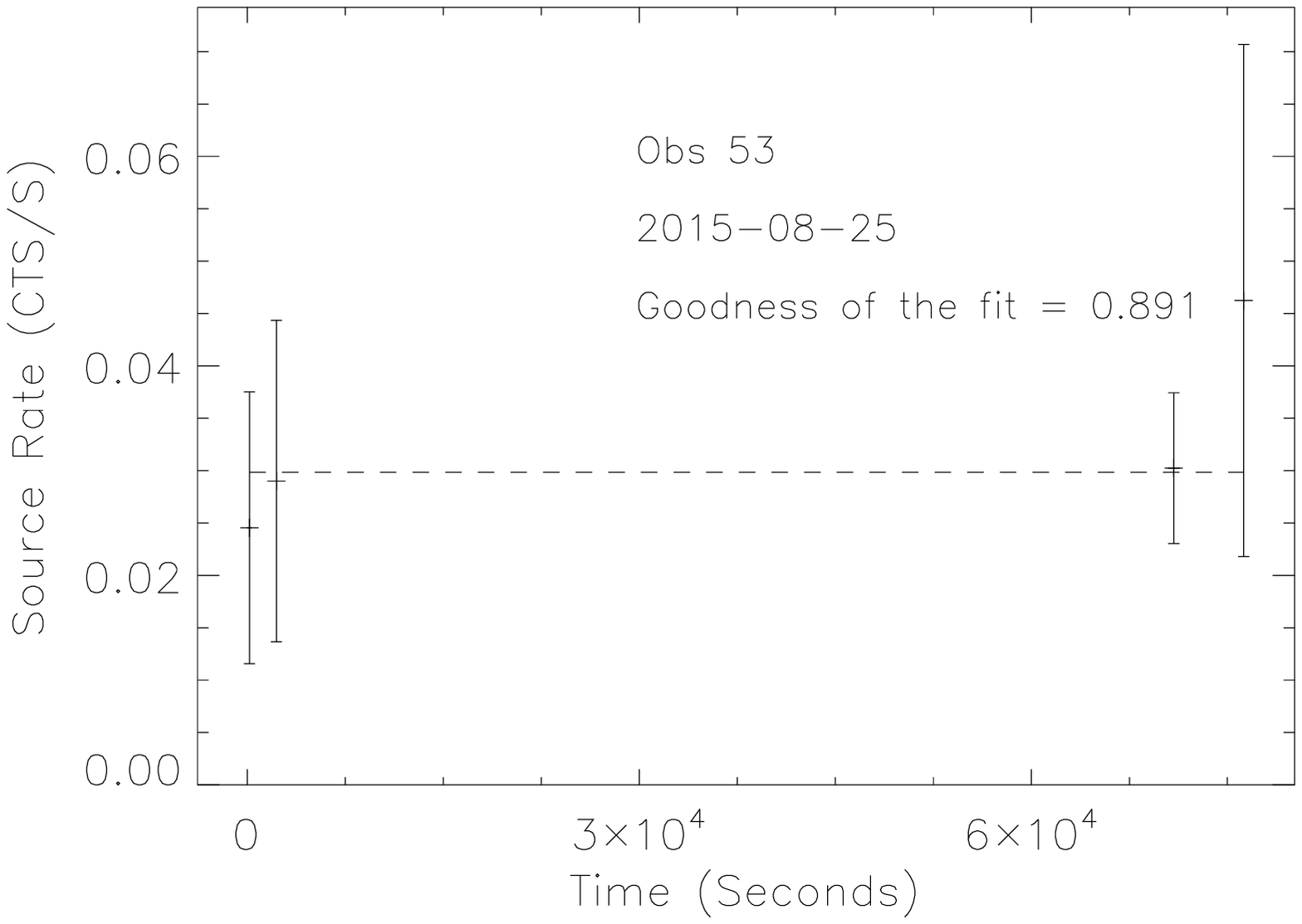}
\includegraphics[width=2.24in, height=1.6in]{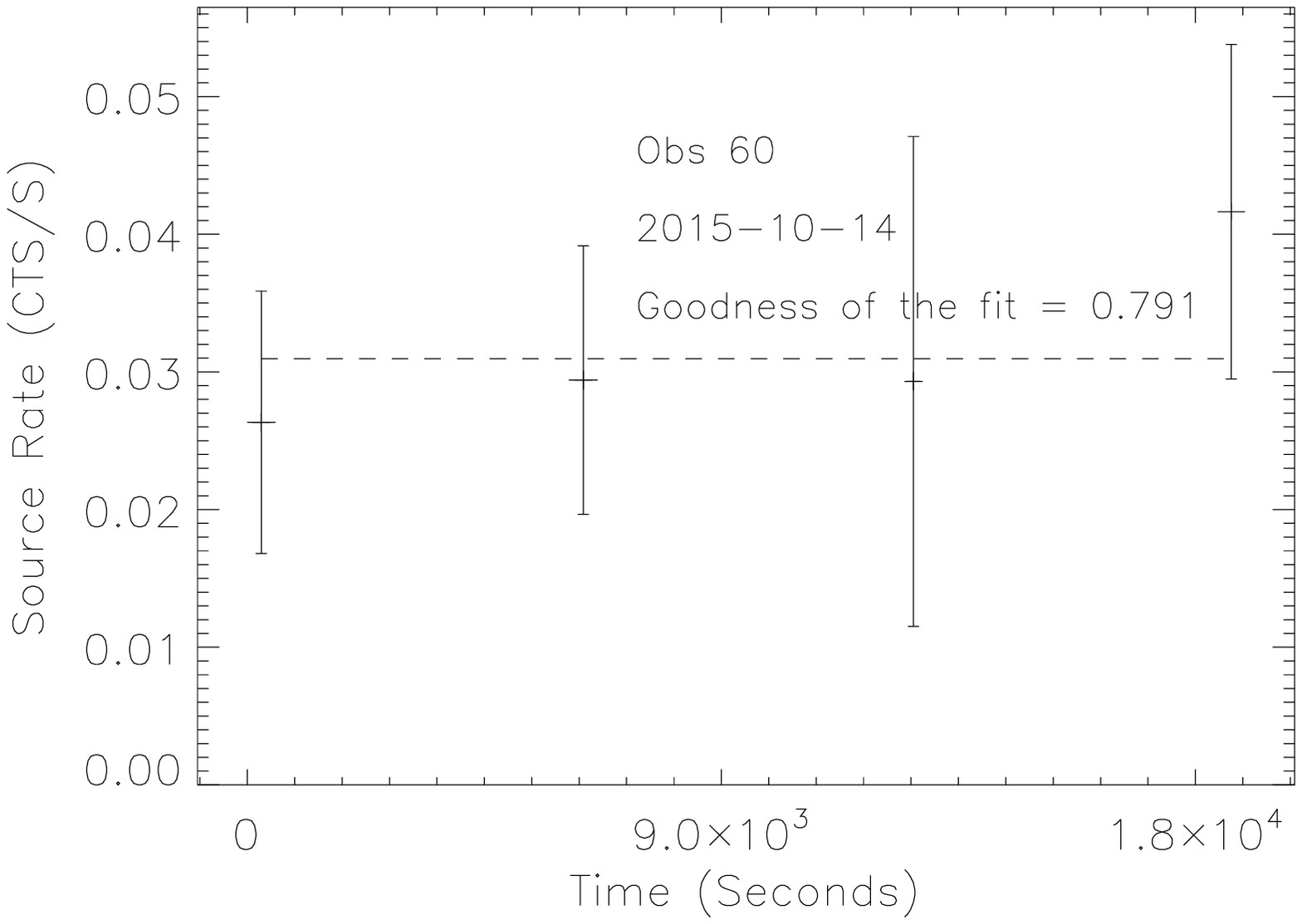}
\includegraphics[width=2.24in, height=1.6in]{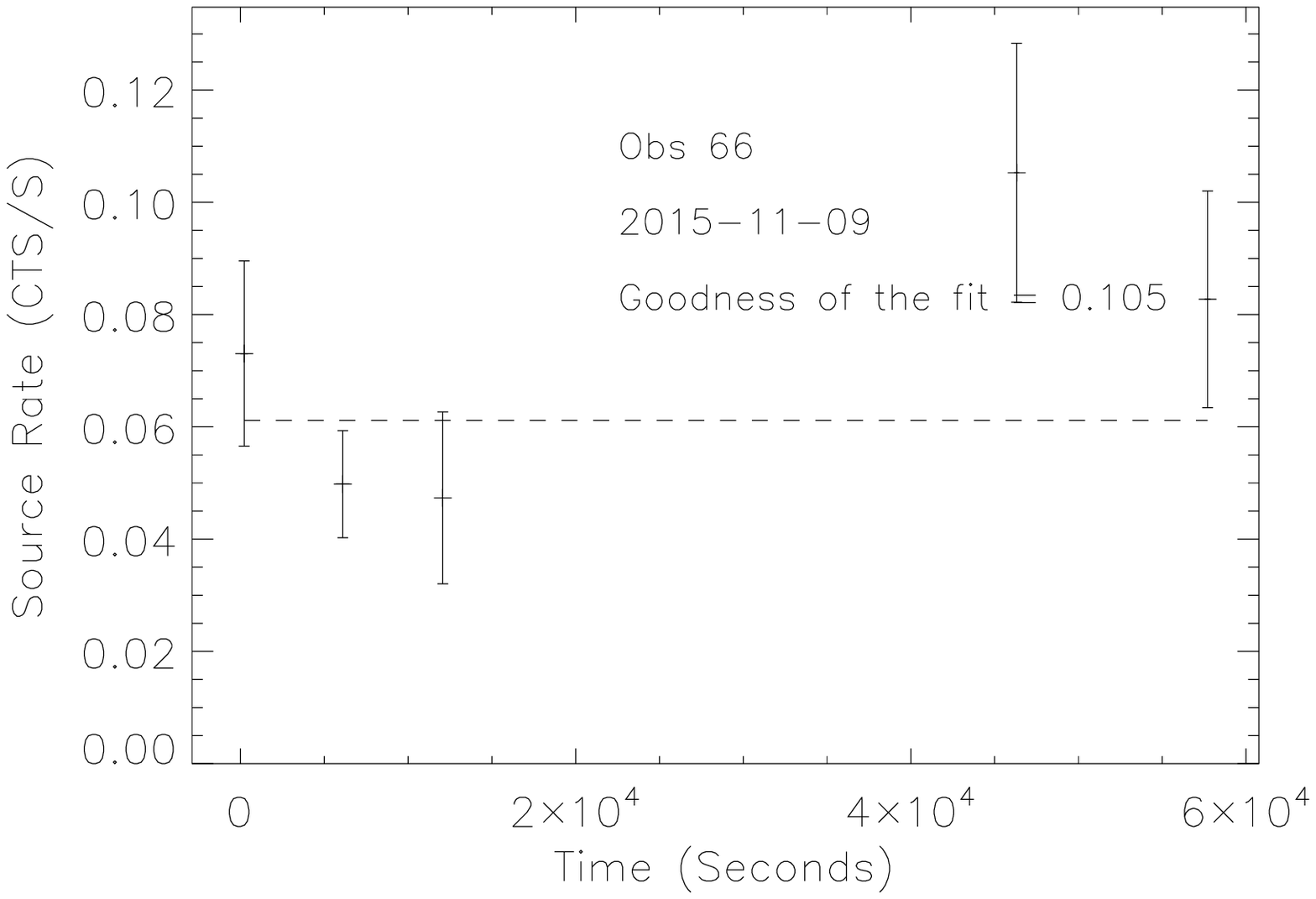}
\includegraphics[width=2.24in, height=1.6in]{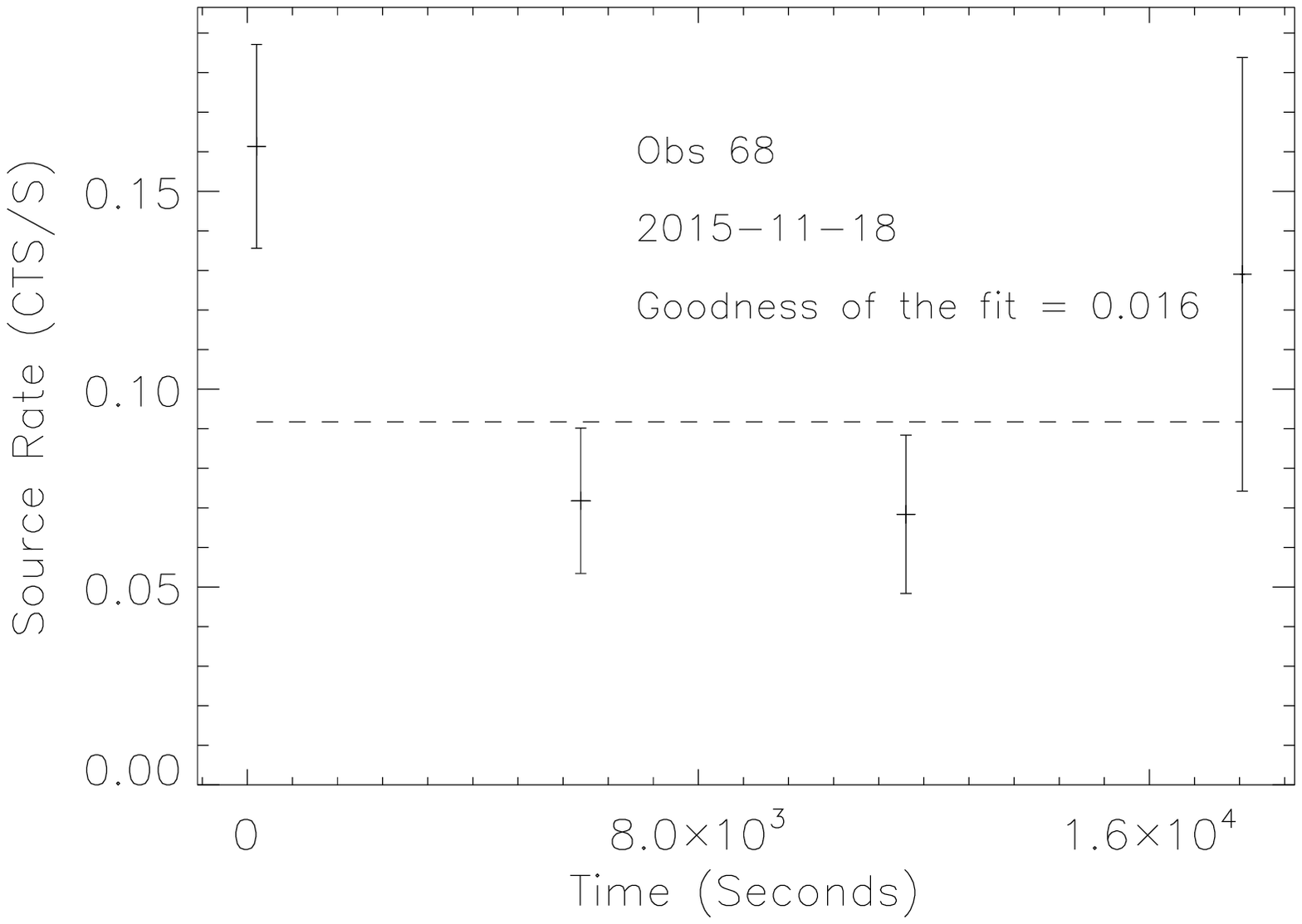}
\includegraphics[width=2.24in, height=1.6in]{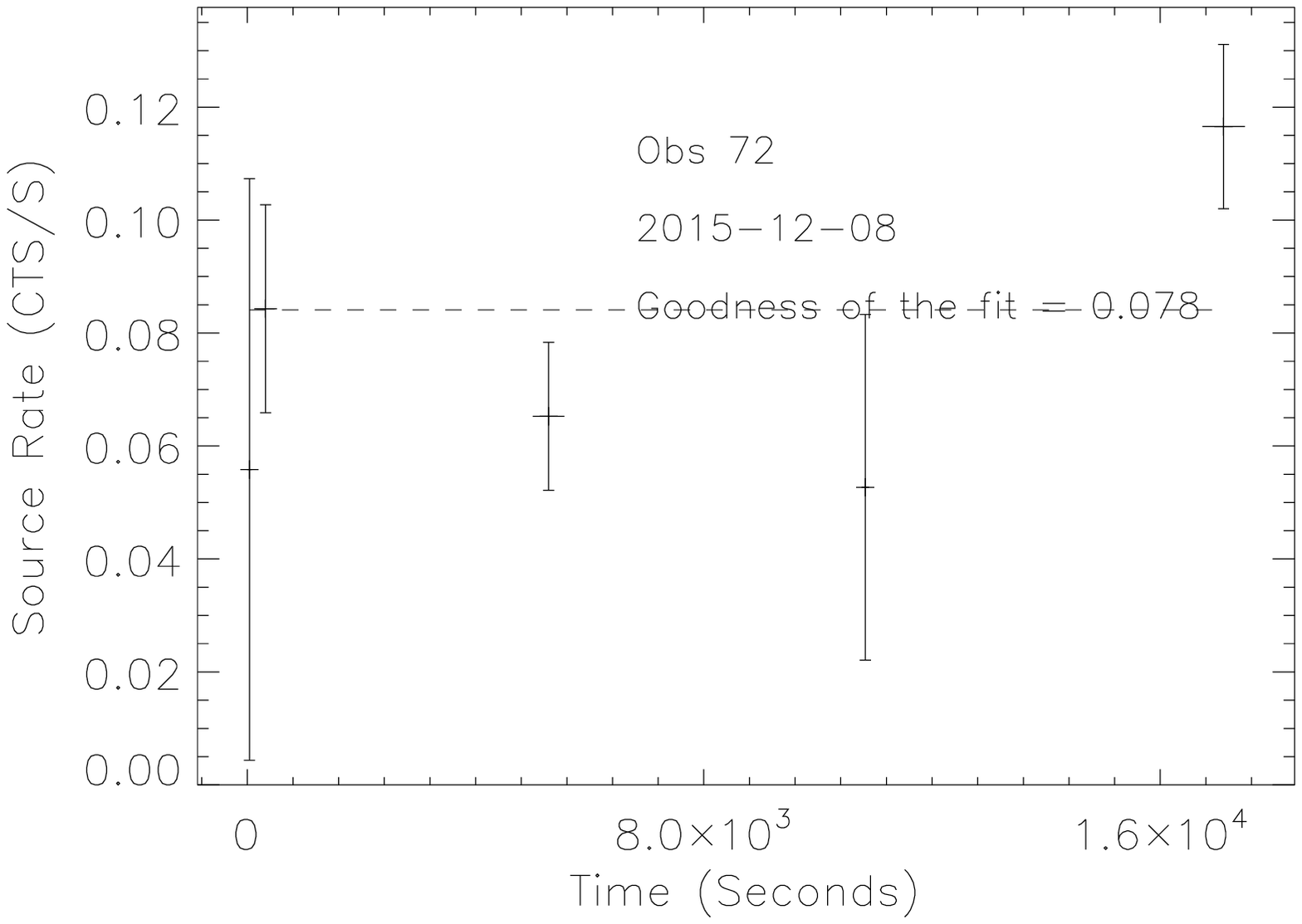}
\includegraphics[width=2.24in, height=1.6in]{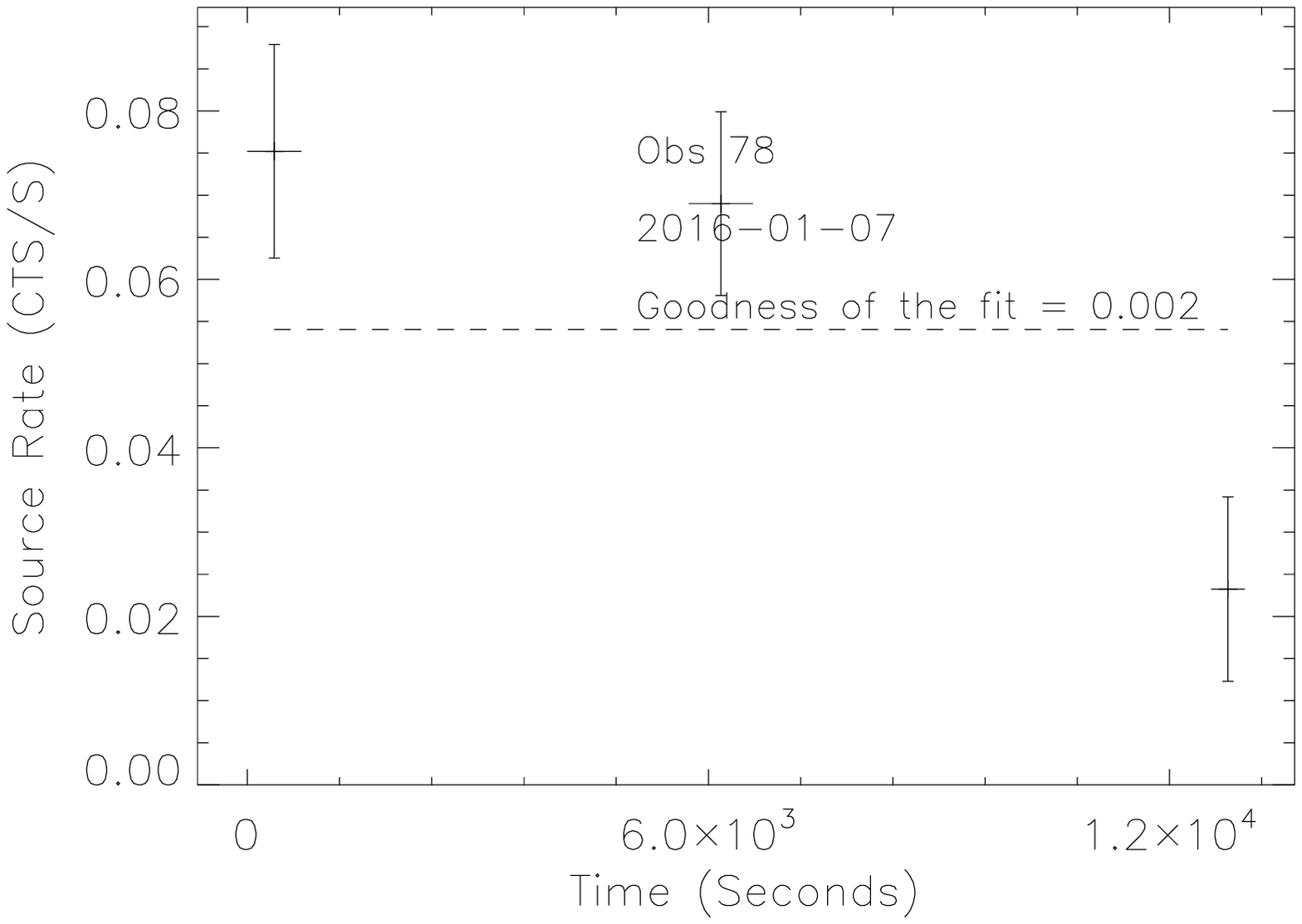}
\end{center}
\caption{The \AG individual light curves (\Rosat and \SwiftE).
The corresponding constant flux level is denoted by a dashed line.
The horizontal bar gives the time bin for each data point.
The observation number corresponds to the last two digits in the
\Swift ObsID (see Section~\ref{sec:data}).
}
\label{fig:dailyLC1}
\end{figure*}

Here, we show a sample of the individual {\it background-subtracted}  
light curves of \AG in the (0.3 - 3 keV) energy range,
based on the good time intervals in the \Rosat and \Swift observations
(see Section~\ref{sec:xray_lc}).
To address the X-ray variability, we fitted a constant count
rate model to each individual LC using $\chi^2$ fitting. The formal
goodness of fit is given in each plot (Fig.~\ref{fig:dailyLC1}).
The 2015 \Swift light curves were chosen to have at least 3 data 
points and to be representative of the short time variability over the
entire observation period (2015 June - 2016 January).

\section{UV short time scale variability of \AG}
\label{append:UV}
\begin{figure*}
\begin{center}
\includegraphics[width=2.24in, height=1.6in]{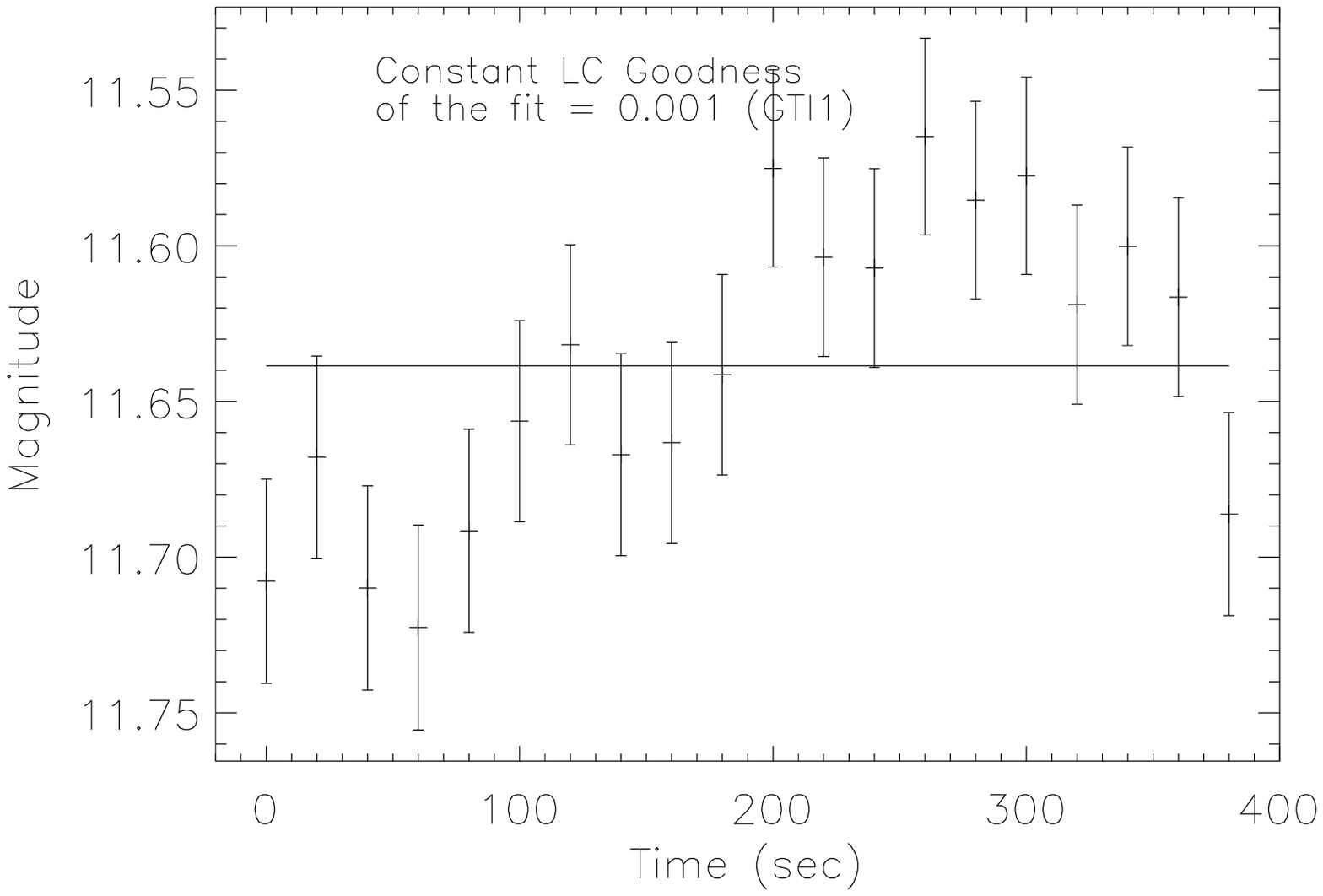}
\includegraphics[width=2.24in, height=1.6in]{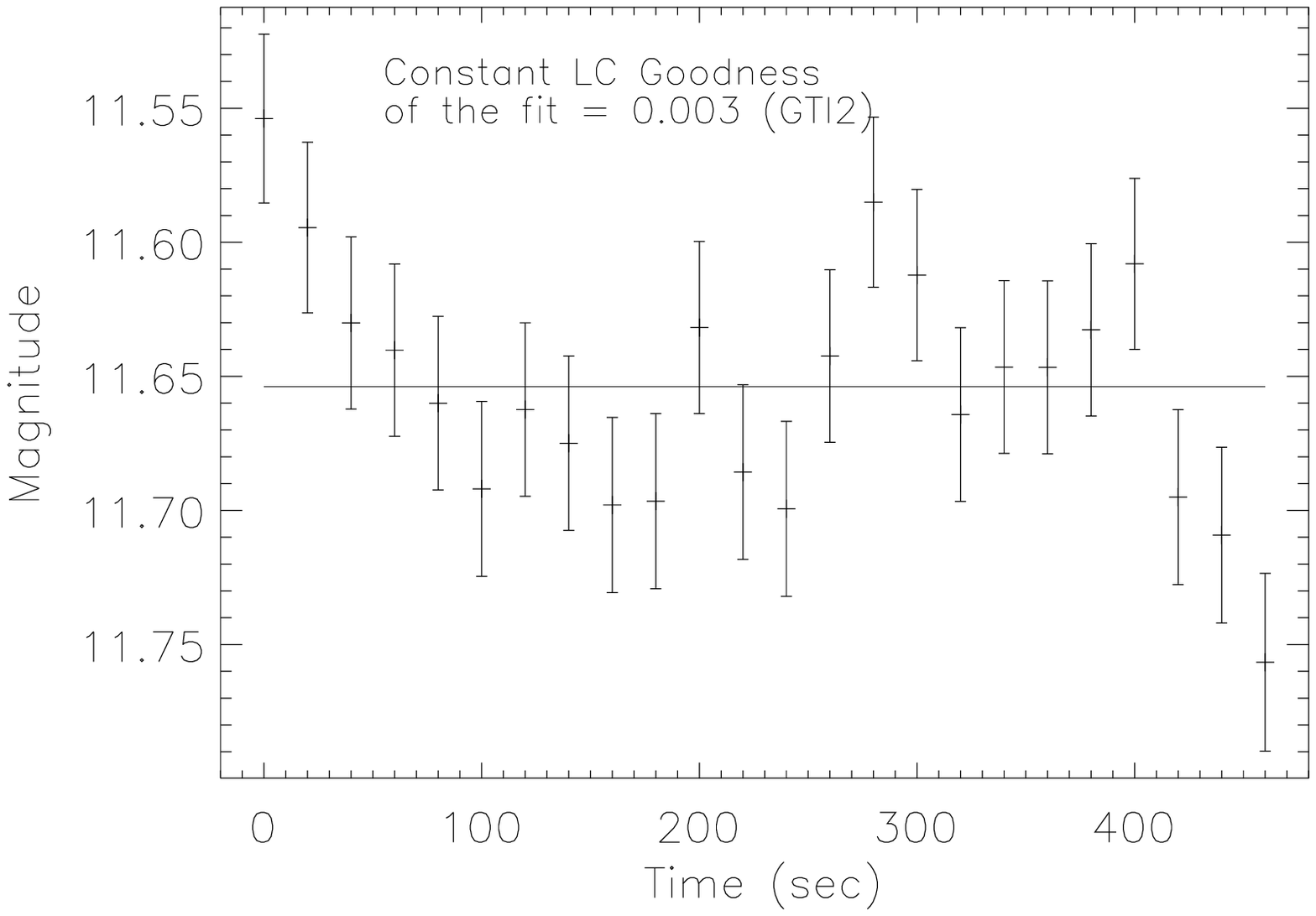}
\includegraphics[width=2.24in, height=1.6in]{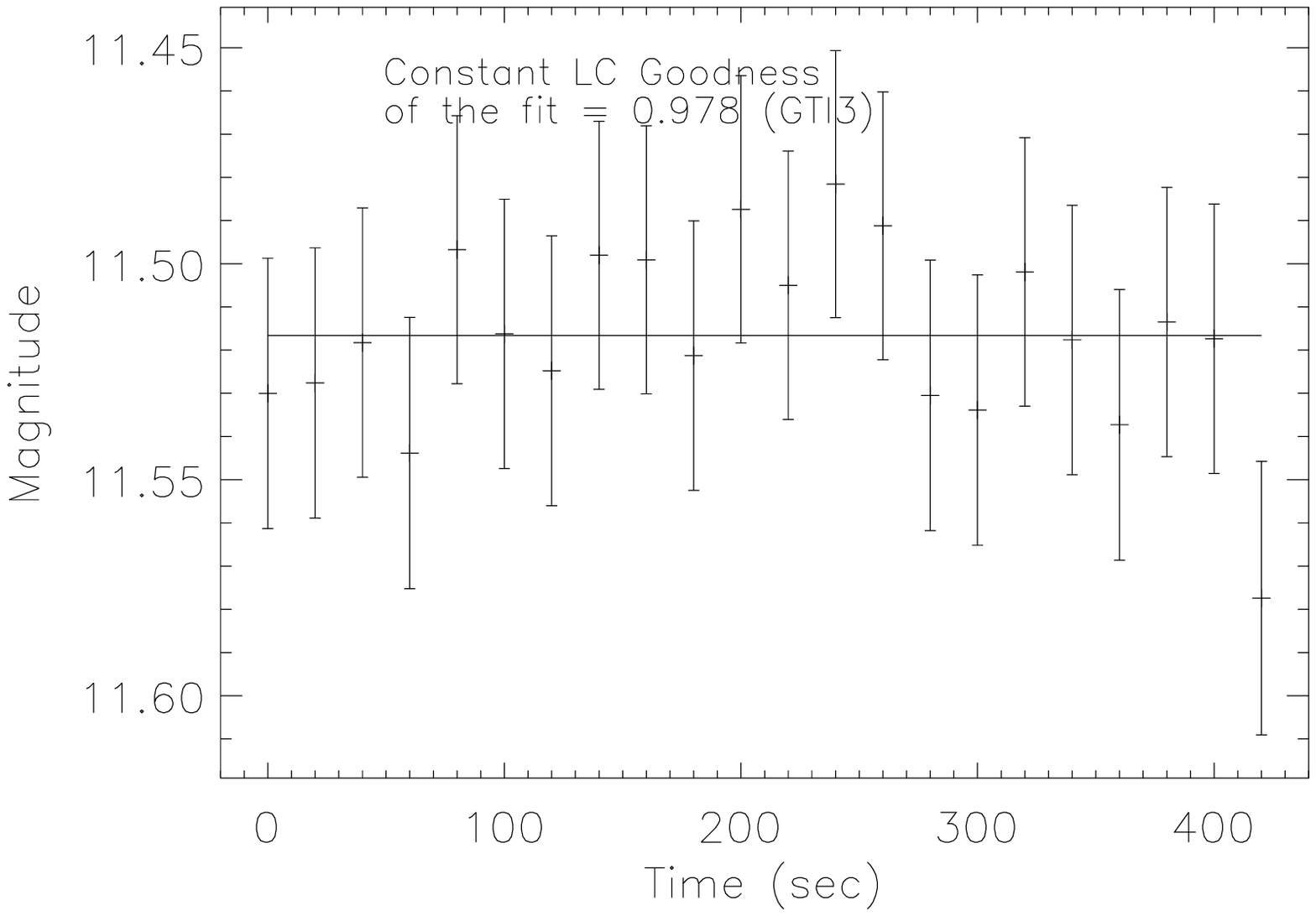}
\includegraphics[width=2.24in, height=1.6in]{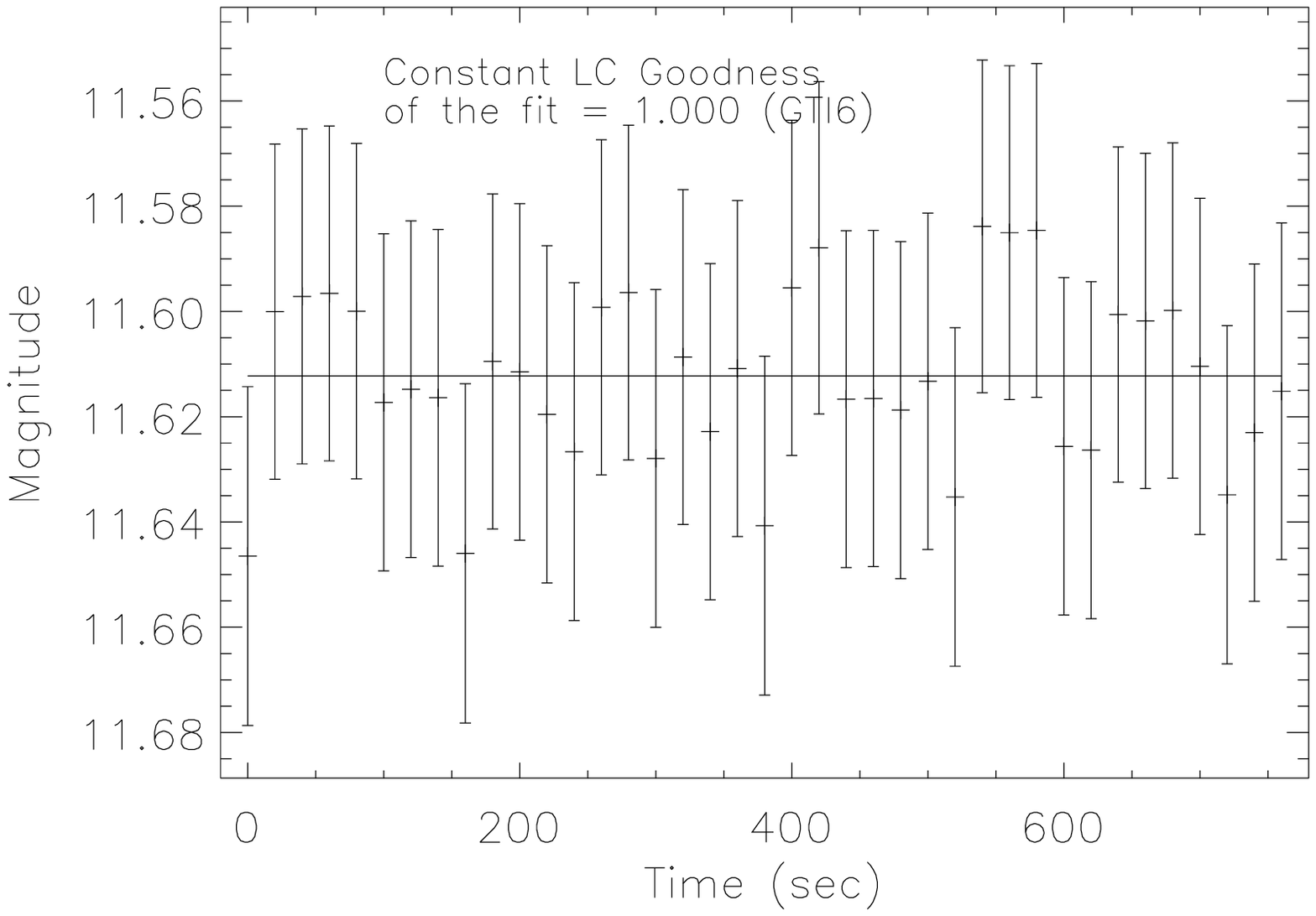}
\includegraphics[width=2.24in, height=1.6in]{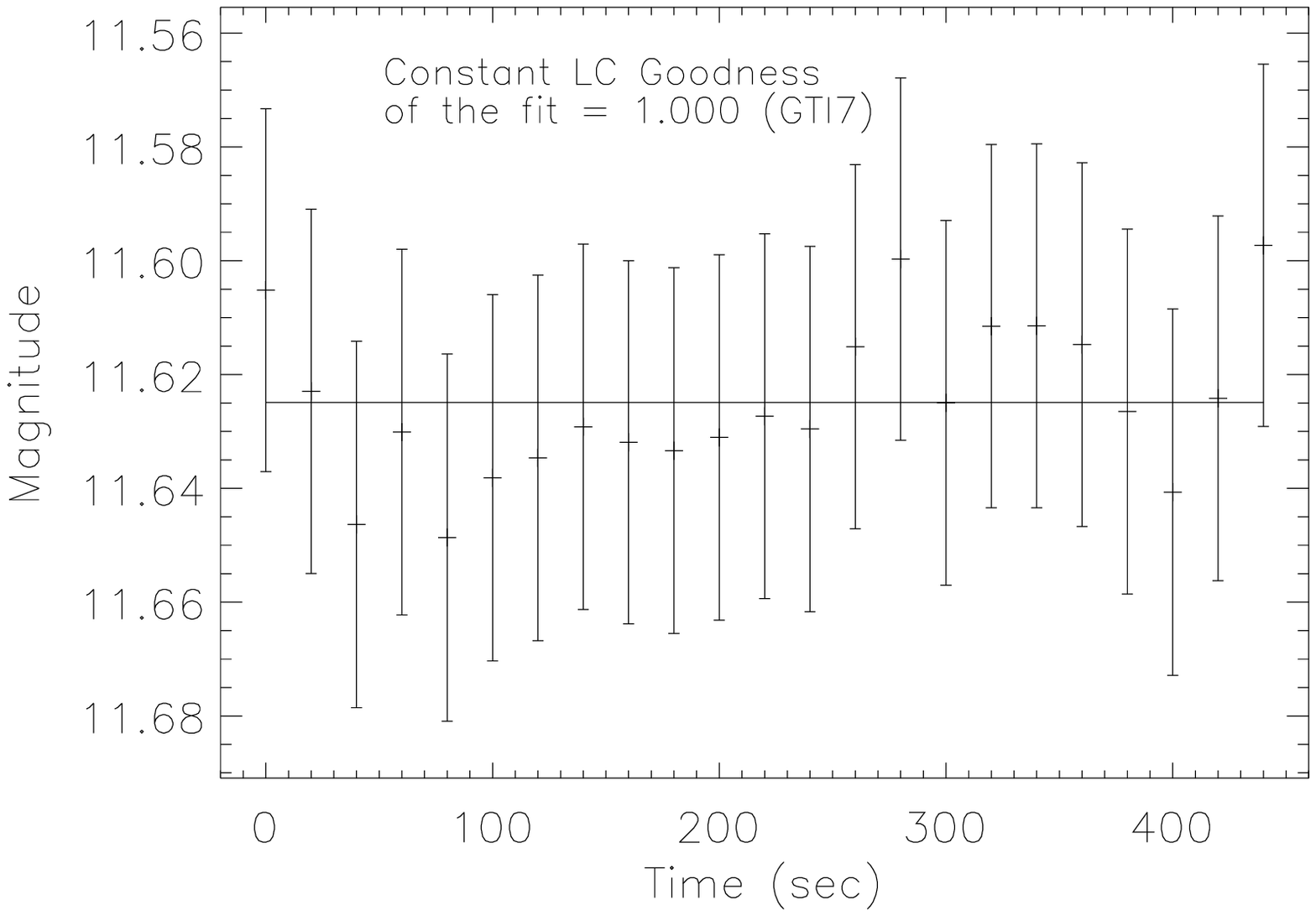}
\includegraphics[width=2.24in, height=1.6in]{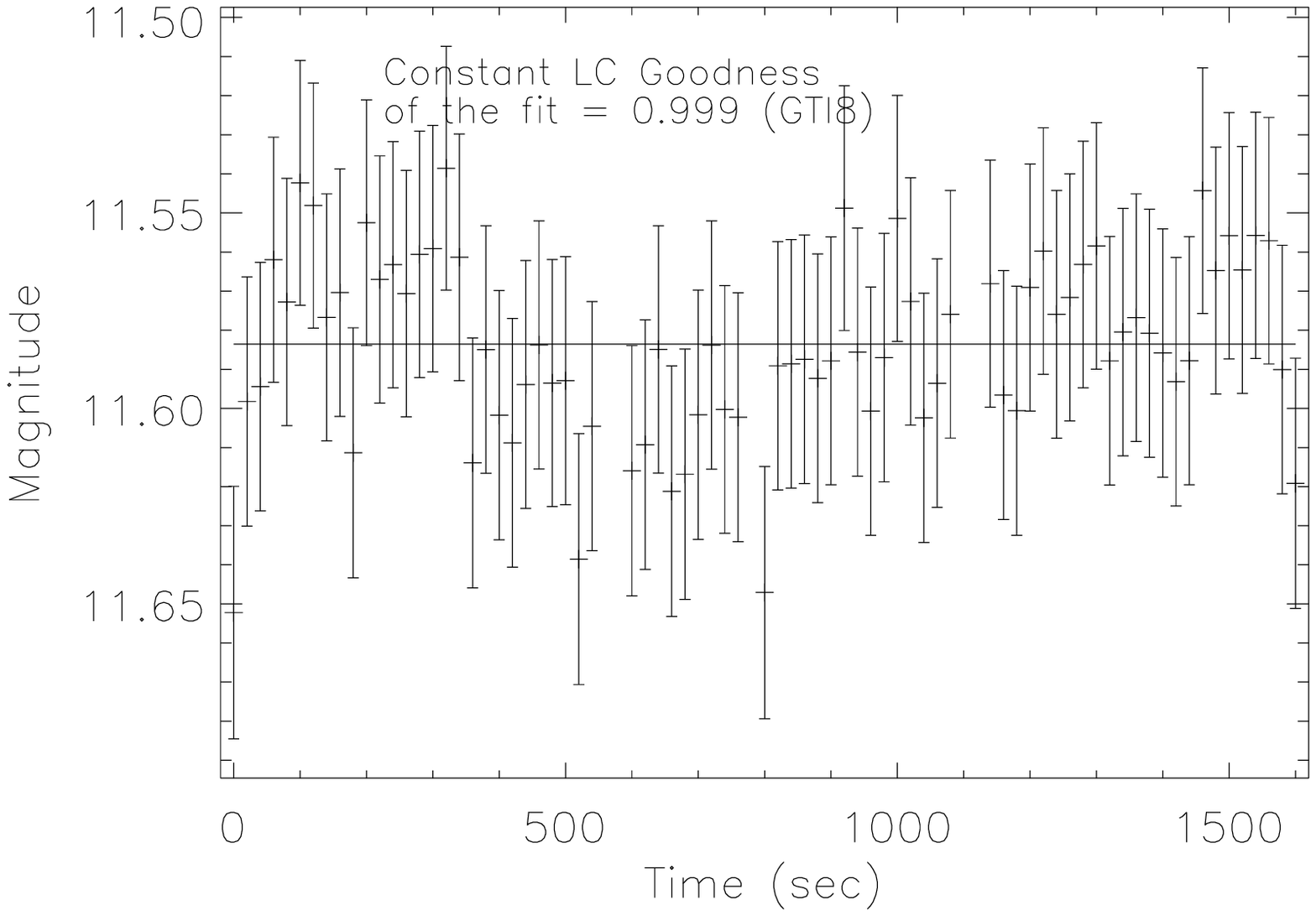}
\includegraphics[width=2.24in, height=1.6in]{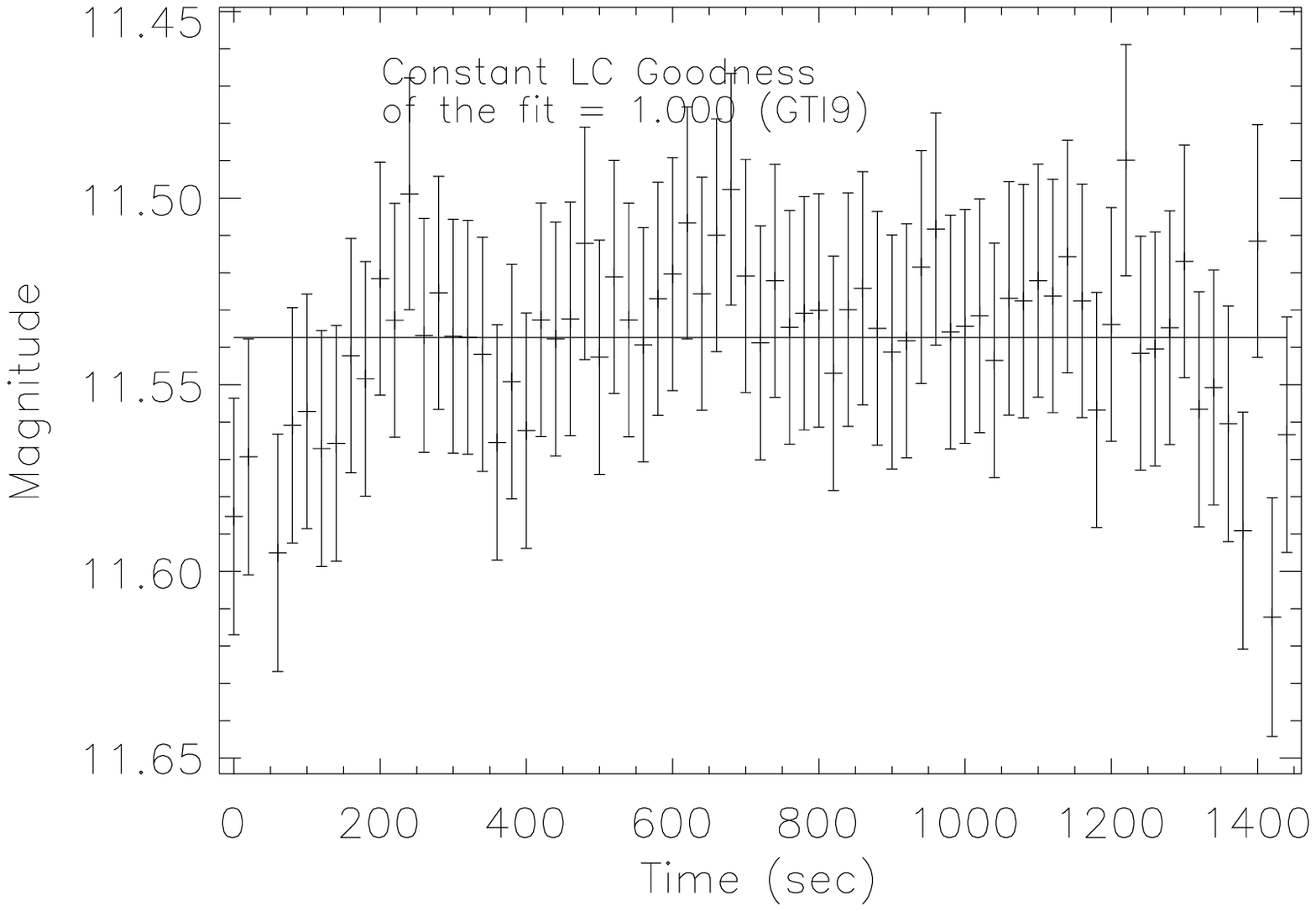}
\includegraphics[width=2.24in, height=1.6in]{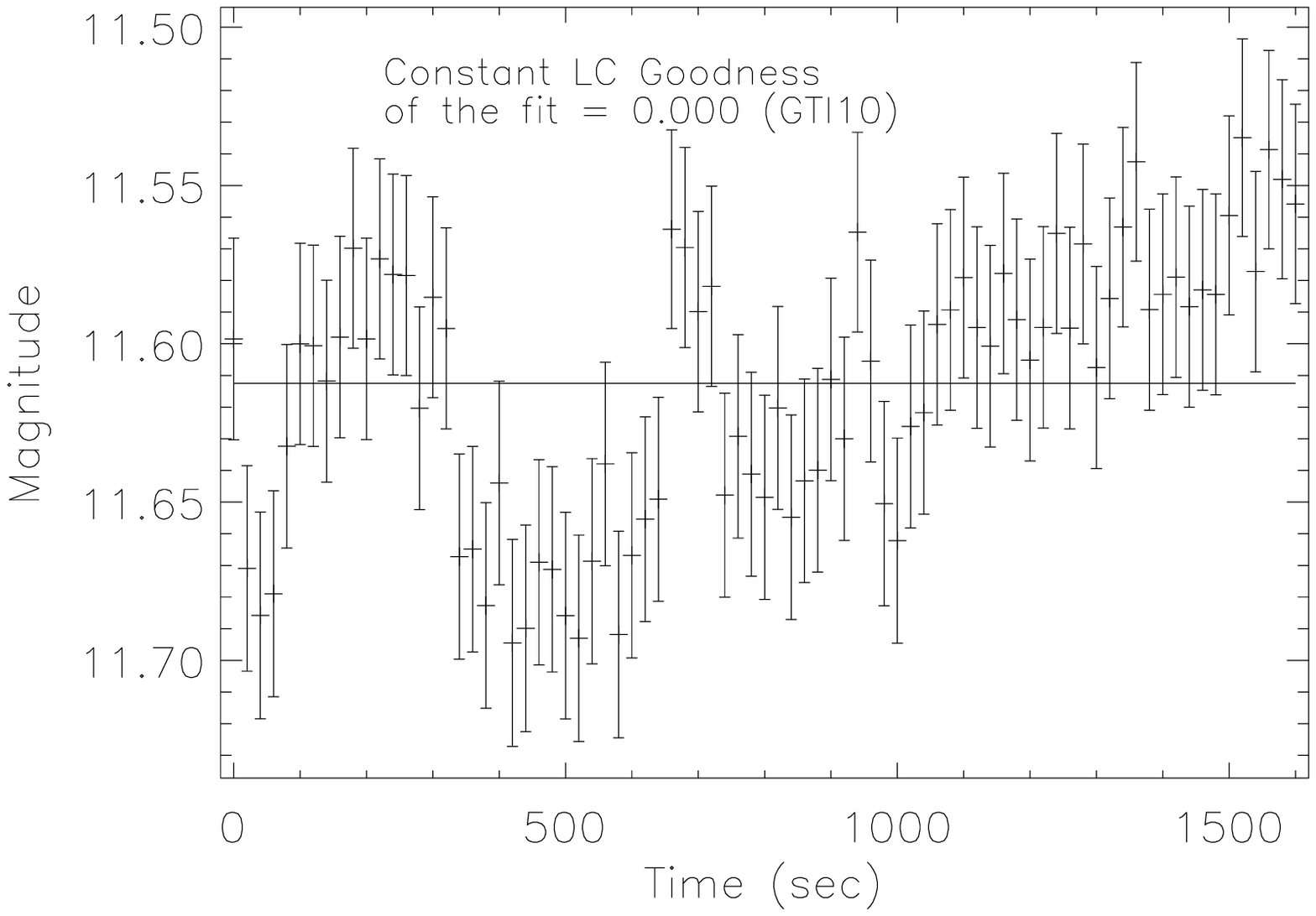}
\includegraphics[width=2.24in, height=1.6in]{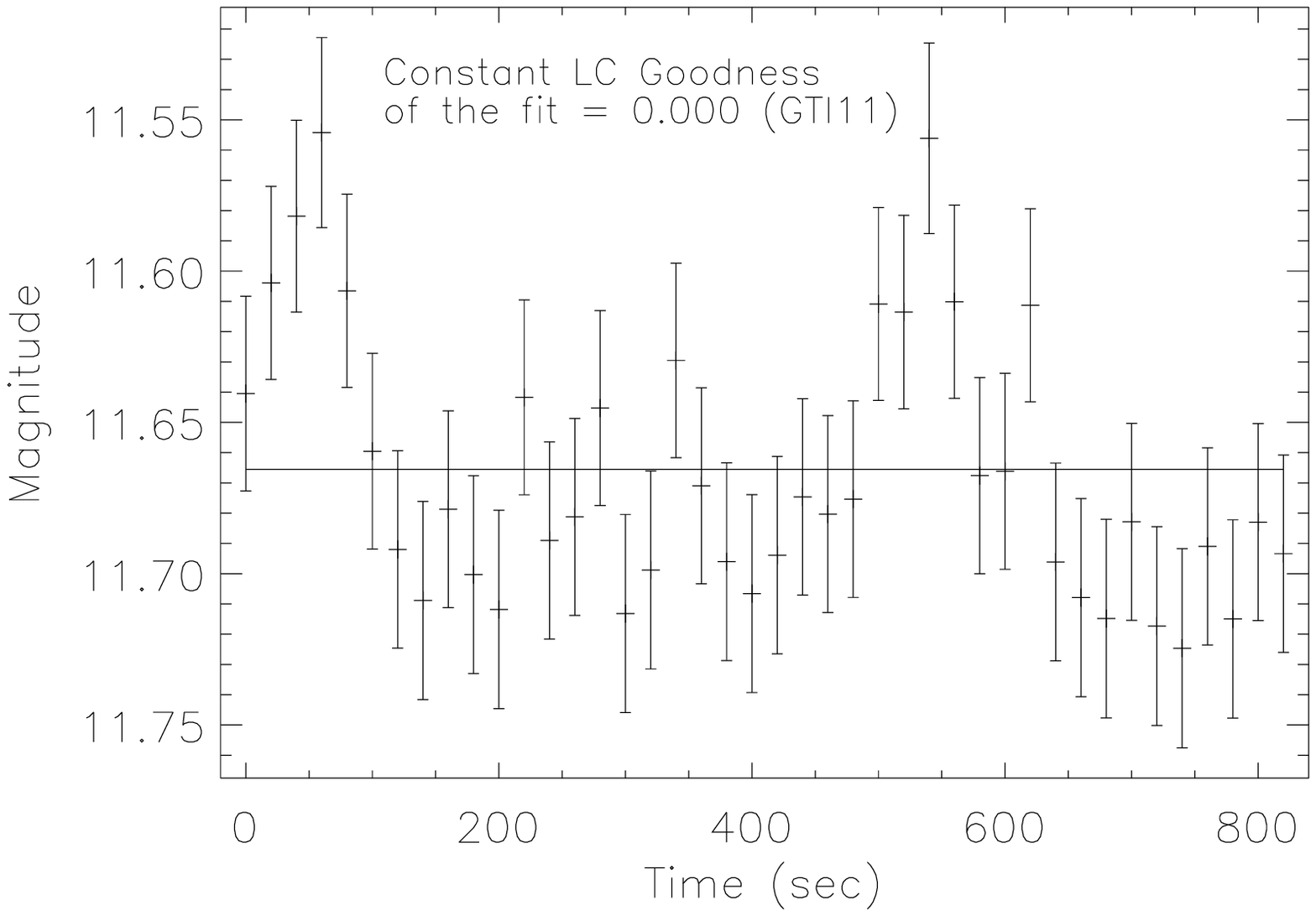}
\includegraphics[width=3.36in, height=2.4in]{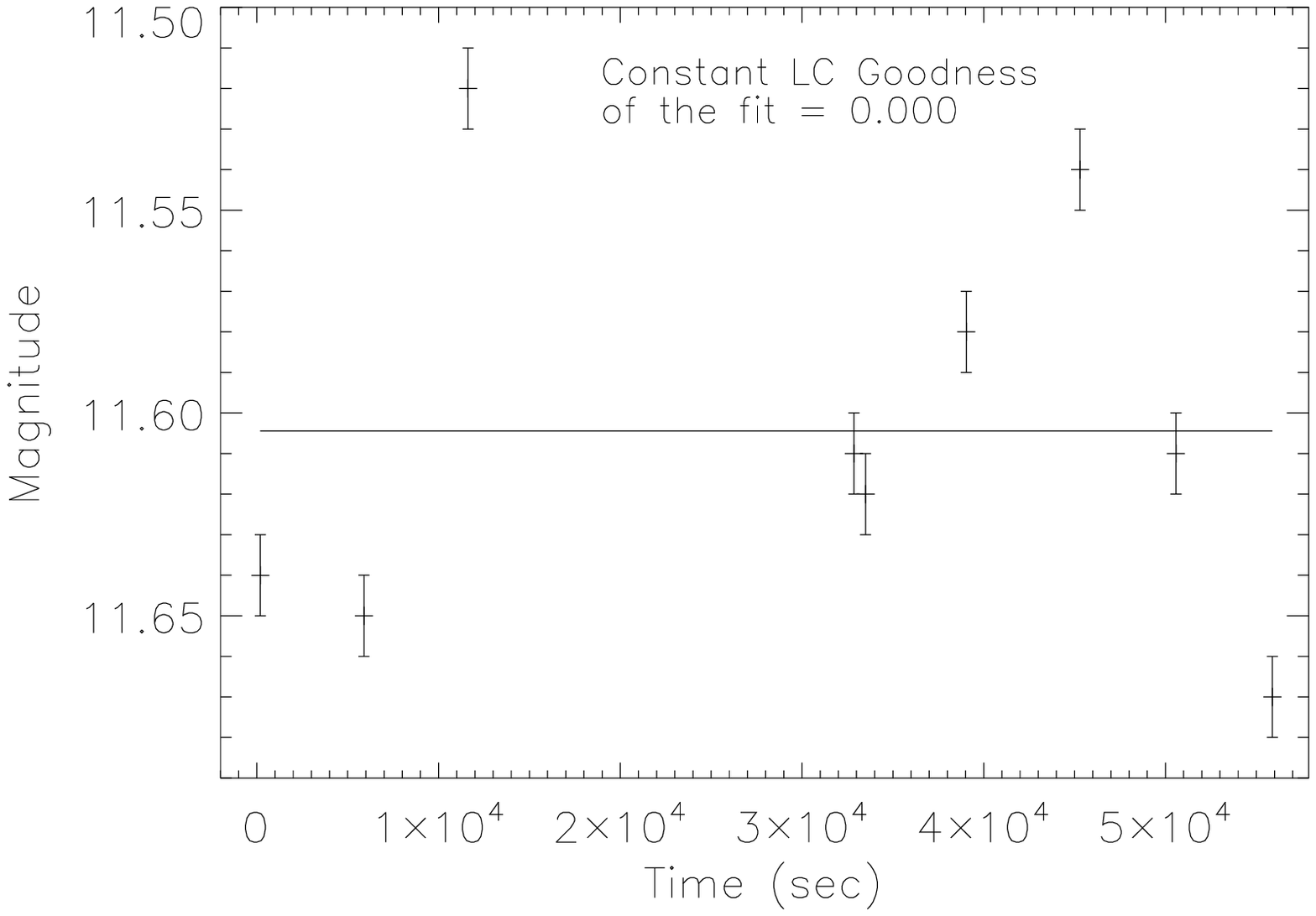}
\end{center}
\caption{The UVOT light curves (time bins of 20 s) of \AG in the
UVM2 filter for
the \Swift observation on 2015 June 28. The corresponding constant
magnitude level is denoted by a dashed line. The consecutive GTI
number is denoted in each panel.
The total LC (GTI-average) is shown in the bottom panel.
}
\label{fig:uvLC}
\end{figure*}

The Ultraviolet/Optical Telescope (UVOT) aboard \Swift allows for
obtaining UV data on a given target simultaneously with its X-ray
data. This was the case for all the \Swift observations of \AG
discussed in this work (see Section~\ref{sec:data}). Here, we show some
results for Obs 3 (ObsID 00032906003) taken on 2015 June 28. On this
date, UVOT observed \AG in the UVM2 filter ($\lambda_0 = 2231$~\AA,
FWHM $= 498$~\AA) taking 11 different exposures (good time intervals;
GTI) in image mode. Following the Swift UVOT Data Reduction
Guide\footnote{\url{http://swift.gsfc.nasa.gov/analysis/UVOT_swguide_v2_2.pdf}},
we analysed the corresponding event data file for each GTI. We
filtered the event file in time domain that allowed us to build the
corresponding short time scale light curves  by making use of the
{\it uvotevtlc} script. We considered only those GTIs with exposures
longer than 300 s (9 in total for Obs 3).  The resultant UV LCs are
shown in Fig.~\ref{fig:uvLC}.
We see that on 2015 June 28
appreciable variability of 0.05 - 0.1 mag was
definitely present in the UV emission of \AG on time scales of minutes
and hours.

\bsp    
\label{lastpage}
\end{document}